\documentclass{aa}  

\usepackage{graphicx}
\usepackage{txfonts}
\usepackage{newtxtext,newtxmath}

\usepackage[T1]{fontenc}
\usepackage{ae,aecompl}

\usepackage{amsmath}	
\usepackage{amssymb}	
\usepackage{lineno,hyperref}
\usepackage{pdflscape}
\usepackage{wasysym}
\usepackage{lscape}

\usepackage[singlelinecheck=off]{caption} 

\usepackage{siunitx}
\newcommand{\percent}{\% }

\begin{document} 

\title{A new method for measuring the meteor mass index: \\
application to the 2018 Draconid meteor shower outburst}

\titlerunning{Novel mass index estimation applied to 2018 Draconids}


\author{D. Vida
    \inst{1, 2}\fnmsep\thanks{\email{dvida@uwo.ca}}
    \and
    M. Campbell-Brown \inst{2, 3}
    \and
    P. G. Brown \inst{2, 3}
    \and
    A. Egal \inst{2}
    \and
    M. J. Mazur \inst{1, 2}
    }

\institute{
    Department of Earth Sciences, University of Western Ontario, London, Ontario, N6A 5B7, Canada
    \and
    Department of Physics and Astronomy, University of Western Ontario, London, Ontario, N6A 3K7, Canada
    \and 
    Centre for Planetary Science and Exploration, University of Western Ontario, London, Ontario, N6A 5B8, Canada
    }

\date{Received 13 December 2019 / Accepted 6 February 2020}

 
  \abstract
   {Several authors predicted an outburst of the Draconid meteor shower in 2018, but with an uncertain level of activity.}
   {Optical meteor observations were used to derive the population and mass indices, flux, and radiant positions of Draconid meteors.}
   {90 minutes of multi-station observations after the predicted peak of activity were performed using highly sensitive Electron Multiplying Charge Coupled Device (EMCCD) cameras. The data calibration is discussed in detail. A novel maximum likelihood estimation method of computing the population and mass index with robust error estimation was developed. We apply the method to observed Draconids and use the values to derive the flux. Meteor trajectories are computed and compared to predicted radiant positions from meteoroid ejection models.}
   {We found that the mass index was $1.74 \pm 0.18$ in the 30 minute bin after the predicted peak, and $2.32 \pm 0.27$ in the next 60 minutes. The location and the dispersion of the radiant matches well to modeled values, but there is an offset of \ang{0.4} in solar longitude.}
   {}


   \maketitle
%



\section{Introduction}

The Draconids are an annual meteor shower whose parent body is the Jupiter-family comet 21P/Giacobini-Zinner. The shower usually has a very low activity with a Zenithal Hourly Rate (ZHR) of $\sim 1$ \citep{jenniskens2006meteor}. It produced large meteor storms in 1933 and 1946 and strong outbursts in a number of other years \citep{egal2019}. Many of these outbursts were not predicted beforehand. The shower's occasionally high intensity and unpredictability have made it a focus of research, particularly as it can pose a significant impact risk to spacecraft in the near-Earth environment \citep{beech1995c, cooke20102011, egal2018draconid}.

In recent years, the shower has produced several notable outbursts. In 2011 an outburst was predicted in advance \citep{watanabe2008activities, maslov2011future, vaubaillon2011coming} and well observed by both radar \citep{kero2012mu, ye2013radar} and optical methods \citep{trigo20132011, borovivcka2014spectral, koten2014three, segon2014draconids, vaubaillon20152011}. That year the outburst reached a ZHR of 350, and had an average mass index of $2.0 \pm 0.1$, but which varied between 1.84 to 2.30 in one hour during the peak \citep{koten2014three}. 

In contrast, the 2012 outburst was not predicted and was only well observed by the Canadian Meteor Orbit Radar (CMOR) \citep{brown2012draconid}. The shower produced a meteor storm at radar sizes (ZHR $\approx 9000 \pm 1000$), but visual observers reported a ZHR almost 2 orders of magnitude lower, ZHR $\sim 200$, suggesting a high mass distribution index, i.e. that the stream was rich in small meteoroids. Unfortunately, due to the bad timing of the peak (maximum over central Asia) and unfavorable weather conditions elsewhere, no optical orbits associated with the outburst were secured. The 2012 outburst was also peculiar in that modeling suggested a very high ($>$ 100 m/s) meteoroid ejection velocity from the parent comet was needed \citep{ye2013unexpected}.

The 2018 outburst was predicted by various authors \citep{kresak1993meteor, maslov2011future, ye2013unexpected, kastinen2017monte, egal2018draconid}, but the predicted activity varied from weak (ZHR 10 - 20) \citep{maslov2011future} to possible meteor storm levels \citep{kastinen2017monte}. The most recent work by \cite{egal2018draconid}, which reproduced well most historic Draconids activity, predicted a peak ZHR of $\sim 80$ at 00:00 UTC on October 9, 2018.

In this work we analyze 1.5 hours of optical observations from Southwestern Ontario just after the peak\footnote{According to visual observations in the IMO database: \url{https://www.imo.net/members/imo_live_shower?shower=DRA&year=2018}} of the 2018 outburst, from 00:00 UTC to 01:30 UTC. We also develop a novel method of population and mass index estimation, and compute these indices using our observations. Finally, we compare model predicted radiants with our multi-station observations and compute the shower flux.

\section{Instruments and observations}

Optical observations were conducted using low-light level video systems in Southwestern Ontario, Canada operated by the University of Western Ontario's Meteor Physics Group. The observations started shortly after local sunset on October 8/9 at around 00:00 UTC and ended at 01:30 UTC due to cloudy weather; thus only a short period after the peak was observed.

The double station observations were made using four Electron Multiplying Charge Coupled Device (EMCCD) N{\"u}v{\"u} HN{\"u} 1024 cameras\footnote{\url{http://www.nuvucameras.com/fr/files/2019/05/NUVUCAMERAS_HNu1024.pdf}} with Nikkor \SI{50}{\milli \metre} f/1.2 lenses. The systems were operated at 32 frames per second, had a limiting stellar magnitude of $+10.0^M$, and a field of view of 15$\times$15 degrees.

Despite clear skies in the region, none of the dozen Southern Ontario Meteor Network (SOMN) all-sky cameras \citep{brown2010development} observed a single Draconid during all of Oct 8-9. These all-sky cameras have an effective limiting meteor magnitude of $-2^M$, qualitatively suggestive of an absence of larger meteoroids in the falling branch of the outburst. The EMCCD cameras did observe many Draconid meteors, which we use in the following analysis. In this section we describe the EMCCD hardware and give details of the data reduction procedure.

\subsection{EMCCD systems}

\begin{figure}
  \includegraphics[width=\linewidth]{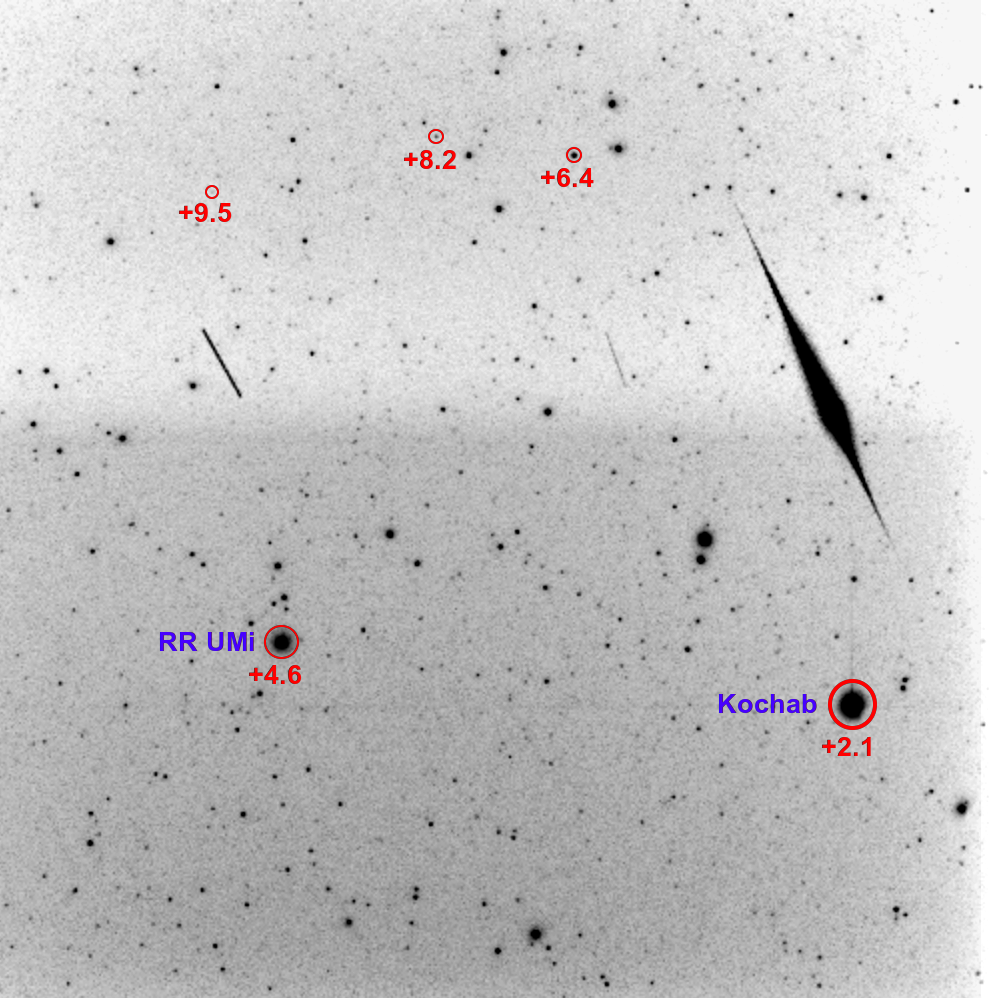}
  \caption{Stack of raw frames showing a $+1^M$ Draconid recorded with camera 01G on October 9, 2018 at 00:01:48 UTC. The meteor was first detected at an apparent magnitude of $+7.0^M$. V-band magnitudes of a selection of stars are shown as well. The vertical gradient of the image background is intrinsic to this specific camera and is compensated for by flat field correction during data reduction. The two thinner and fainter lines left of the meteor are satellites.}
  \label{fig:emccd_meteor}
\end{figure}

\begin{figure}
  \includegraphics[width=\linewidth]{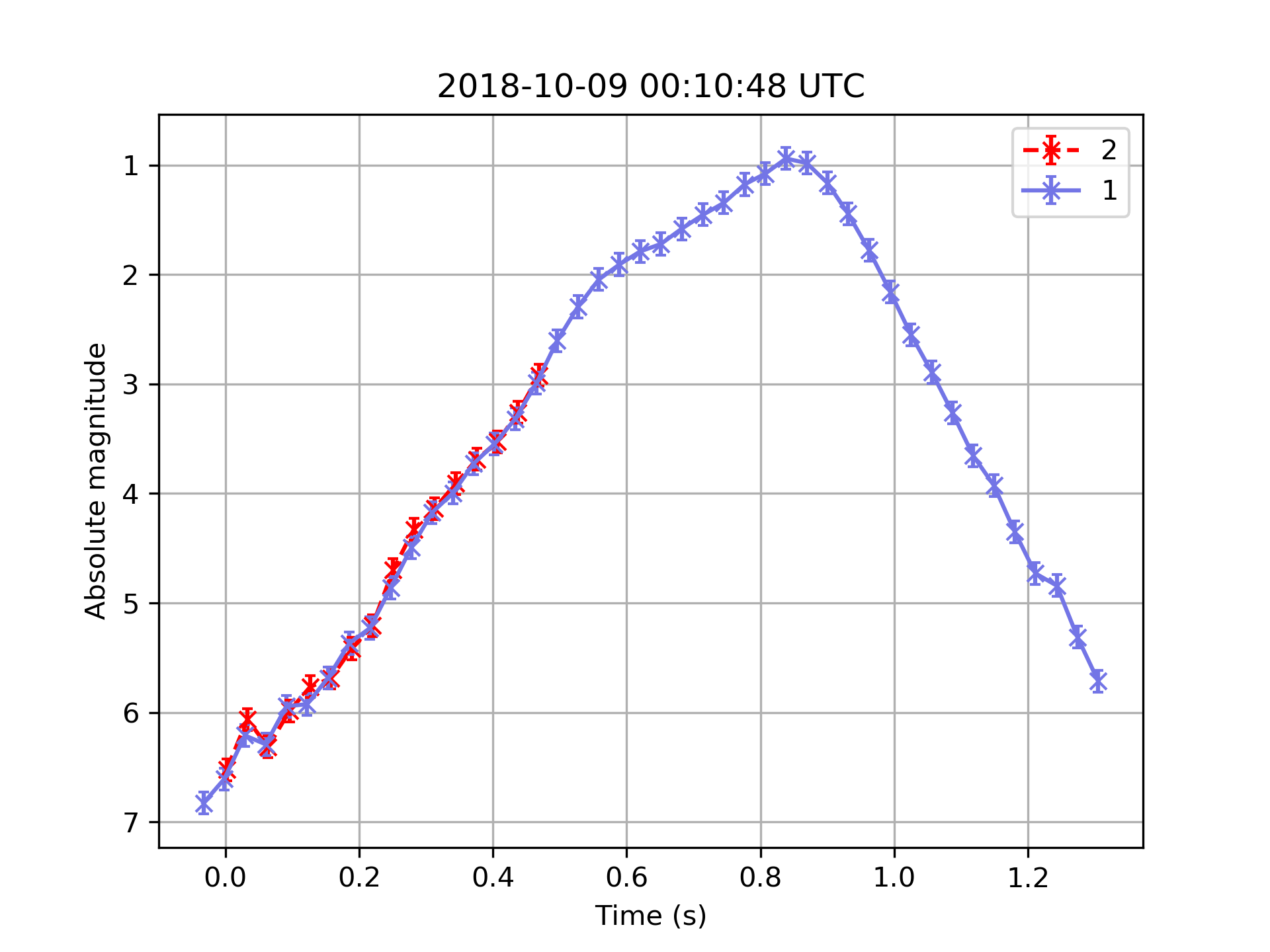}
  \caption{Absolute GAIA G-band magnitude light curve of the meteor in Figure \ref{fig:emccd_meteor}. The error bars are $\sim 0.1$ magnitude. Observations from camera 01G (Tavistock) are shown in blue, and observations from camera 02G (Elginfield) are shown in red.}
  \label{fig:emccd_meteor_lightcurve}
\end{figure}

The N{\"u}v{\"u} HN{\"u} 1024 EMCCD cameras are the most recent addition to the Canadian Automated Meteor Observatory (CAMO) systems \citep{weryk2013canadian}. These 16-bit cameras use electron multiplying technology to increase the number of electrons accumulated in the registers before they are amplified, digitized and read out, which greatly increases the signal-to-noise ratio. In combination with a fast \SI{50}{\milli \metre} f/1.2 lens, the single frame limiting magnitude at 32 frames per second for stars is around $+10.0^M$, and $+7.5^M$ for meteors. The native resolution of the camera is $1024 \times 1024$ pixels, but the frames are binned to $512 \times 512$ to make possible the higher frame rate, which provides for better trajectory precision. The field of view is $\ang{14.75} \times \ang{14.75}$, and the pixel scale is 1.7 arcmin/pixel. All camera were run at a gain setting of 200. These systems are more sensitive and have less noise than the previous generation of systems which relied on Gen III image intensifiers \citep{campbell2015population}. The cameras each have external GPS timing directly encoded at the frame acquisition stage. Figure \ref{fig:emccd_meteor} shows a sample meteor captured with the EMCCD camera and Figure \ref{fig:emccd_meteor_lightcurve} shows the light curve of the same meteor. Note the agreement between the sites in range-corrected magnitudes (normalized to \SI{100}{\kilo \metre}). In Appendix \ref{appendix:lightcurves} we give more examples of multi-station Draconid light curves observed by the EMCCD systems on Oct 9, 2018.

Two cameras are deployed at the each of the two CAMO sites in Southwestern Ontario. Cameras 01F and 01G are located at Tavistock (\ang{43.26420} N, \ang{80.77209} W, \SI{329}{\metre}), and 02F and 02G at Elginfield (\ang{43.19279} N, \ang{81.31565} W, \SI{324}{\metre}). The distance between the stations is about \SI{45}{\kilo \metre}. Both pairs are pointed roughly north, one pair (the G cameras) at the elevation of 45 degrees, the other (F cameras) at 65 degrees. There was no volume overlap between the F and G cameras. One camera from each site had the Draconid radiant inside the field of view. The EMCCDs captured a total of 92 double station Draconids ranging in peak magnitudes from $+1^M$ to $+6.5^M$, of which only 68 had complete light curves from at least one camera.

\subsection{Data reduction and calibration}

Using the tools available in the open-source RMS (Raspberry pi Meteor Station) library\footnote{RMS GitHub web page: \url{https://github.com/CroatianMeteorNetwork/RMS}}, automated meteor detection \citep[described in detail in][]{vida2016open} was run on all collected data and both the astrometric and photometric calibration was performed manually. The automated detections were manually classified and only Draconids were extracted for further analysis. 

\begin{figure}
  \includegraphics[width=\linewidth]{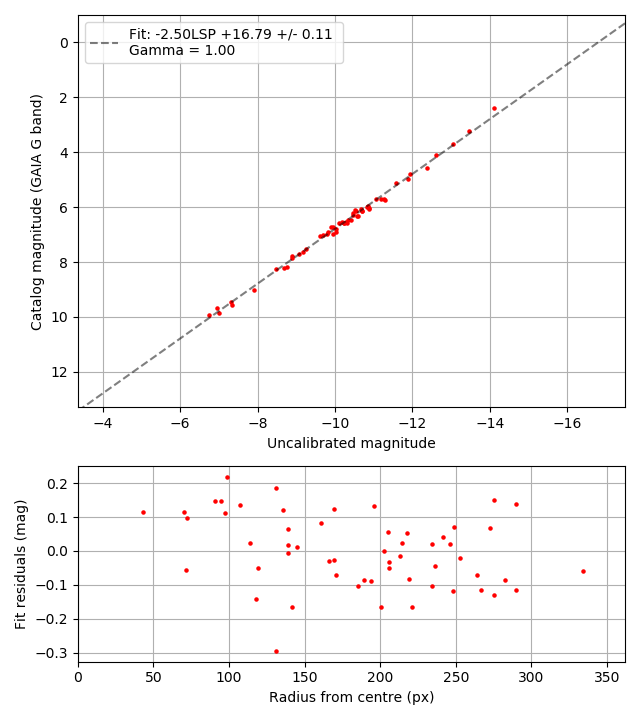}
  \caption{Photometric calibration of the 01F camera. The uncalibrated magnitude is simply $-2.5 \log_{10} I_s$, where $I_s$ is the background-subtracted sum of pixel intensities of all star pixels (LSP in the figure). The sensor response was linear (Gamma = 1.0). The magnitudes of stars used for calibration ranged from $+2.3^M$ to $9.9^M$.}
  \label{fig:01F_photometry}
\end{figure}

The astrometric and photometric calibrations were done using the Gaia DR2 star catalog \citep{gaia2018dr2} as the GAIA G spectral band matches the spectral response of the EMCCDs well. We applied flat fields and bias frames to the video data using the method described in \cite{berry2000astronomical}.
The average residuals of the astrometric calibration were $\sim 1/10$ pixels or $\sim 10$ arc seconds, and the accuracy of the photometric calibration was $\pm 0.1$ magnitude. Figure \ref{fig:01F_photometry} shows the photometric calibration, and Table \ref{tab:01F_astrometry_report} details the astrometric calibration for the 01F EMCCD camera.

\begin{table*}[t] 
	\caption{Example astrometric calibration for the 01F EMCCD camera. The estimated field of view was \ang{14.75} $\times$ \ang{14.75}. The average error was 0.08 px and 0.15 arc minutes. Img X and Y are star image coordinates on the chip, RA and Dec are equatorial coordinates in J2000, Mag is the GAIA G band magnitude, -2.5*LSP is the uncalibrated magnitude (LSP is the logarithm of the sum of pixel intensities), Cat X and Y are predicted catalog star positions in image coordinates, error columns are fit residuals in arc minutes and pixel, and $\theta$ is the angle of a vector pointing from the predicted to the measured star position. The photometric offset was $+16.80 \pm 0.11$, but some stars used for photometry were not used for astrometry due to their smaller signal to noise ratio.}
	\label{tab:01F_astrometry_report} 
	\centering 
	
	\begin{tabular}{l r r r r r r r r r r r} 
	\hline\hline 
No &   Img X &   Img Y & RA & Dec &    Mag & -2.5*LSP &   Cat X &   Cat Y & Error  &  Error & Error $\theta$ \\
 &   (px) &   (px) & (deg) & (deg) &    & &   (px) &   (px) & (arcmin) & (px) & (deg) \\
\hline
  1 &  234.34 &  163.91 &  267.193 &   +48.836 &  +6.91 &   -10.02 &  234.50 &  163.97 &     0.29 &    0.17 &     +21.8 \\
  2 &  370.26 &   15.08 &  272.091 &   +44.113 &  +6.72 &    -9.95 &  370.31 &   15.06 &     0.10 &    0.06 &     -21.2 \\
  3 &  417.60 &  200.09 &  275.386 &   +49.122 &  +4.10 &   -12.62 &  417.56 &  200.10 &     0.16 &    0.04 &    +158.7 \\
  4 &  414.03 &  481.25 &  278.144 &   +57.046 &  +4.56 &   -12.37 &  414.02 &  481.28 &     0.02 &    0.03 &    +101.4 \\
  5 &  187.10 &   62.08 &  264.866 &   +46.006 &  +3.70 &   -13.03 &  187.16 &   61.96 &     0.22 &    0.13 &     -64.1 \\
  6 &  407.94 &   59.93 &  273.885 &   +45.209 &  +6.13 &   -10.53 &  407.93 &   59.93 &     0.11 &    0.02 &    +178.2 \\
  7 &  123.98 &  472.02 &  262.682 &   +57.877 &  +5.72 &   -11.05 &  124.05 &  472.06 &     0.12 &    0.08 &     +31.2 \\
  8 &  100.16 &  294.76 &  261.350 &   +52.790 &  +6.30 &   -10.57 &  100.17 &  294.77 &     0.02 &    0.01 &     +26.3 \\
  9 &  326.99 &  244.11 &  271.723 &   +50.823 &  +6.02 &   -10.87 &  326.98 &  244.12 &     0.10 &    0.02 &    +123.6 \\
 10 &  393.84 &  274.09 &  274.983 &   +51.348 &  +5.99 &   -10.82 &  393.82 &  274.09 &     0.12 &    0.02 &    +166.3 \\
 11 &  130.40 &  107.98 &  262.570 &   +47.403 &  +7.01 &    -9.69 &  130.46 &  108.11 &     0.28 &    0.15 &     +64.0 \\
 12 &  262.78 &  327.09 &  269.245 &   +53.444 &  +8.25 &    -8.48 &  262.84 &  327.19 &     0.34 &    0.11 &     +56.8 \\
 13 &  185.98 &  263.89 &  265.341 &   +51.818 &  +5.70 &   -11.20 &  185.95 &  263.97 &     0.09 &    0.08 &    +113.6 \\
 14 &  344.09 &  278.87 &  272.756 &   +51.740 &  +6.99 &    -9.94 &  344.17 &  278.95 &     0.31 &    0.11 &     +45.8 \\
 15 &  260.64 &   89.92 &  268.004 &   +46.643 &  +6.08 &   -10.68 &  260.43 &   89.93 &     0.35 &    0.21 &    +176.2 \\
 16 &  372.69 &  231.22 &  273.677 &   +50.244 &  +7.85 &    -8.89 &  372.67 &  231.23 &     0.12 &    0.03 &    +157.8 \\
 17 &   92.56 &  381.21 &  260.991 &   +55.279 &  +7.69 &    -9.06 &   92.56 &  381.26 &     0.05 &    0.05 &     +80.9 \\
 18 &  181.62 &  192.91 &  264.950 &   +49.780 &  +6.58 &   -10.32 &  181.58 &  192.95 &     0.11 &    0.07 &    +136.5 \\
 19 &  289.11 &  467.85 &  271.500 &   +57.360 &  +6.58 &   -10.23 &  289.12 &  467.85 &     0.14 &    0.00 &     +44.8 \\
 20 &   89.87 &  441.60 &  260.844 &   +57.012 &  +6.46 &   -10.40 &   89.85 &  441.63 &     0.02 &    0.04 &    +118.2 \\
 21 &  452.50 &  314.07 &  278.047 &   +52.116 &  +6.55 &   -10.29 &  452.48 &  314.18 &     0.23 &    0.11 &    +105.6 \\
 22 &  485.15 &  102.90 &  277.345 &   +45.982 &  +7.04 &    -9.62 &  485.25 &  102.86 &     0.19 &    0.10 &     -21.0 \\
 23 &  491.59 &  327.00 &  279.970 &   +52.196 &  +5.97 &   -10.84 &  491.55 &  327.01 &     0.09 &    0.04 &    +167.3 \\
 24 &   21.27 &  282.99 &  257.628 &   +52.409 &  +6.27 &   -10.47 &   21.23 &  282.97 &     0.11 &    0.05 &    -153.8 \\
 25 &   23.80 &  190.04 &  257.918 &   +49.746 &  +6.14 &   -10.69 &   23.79 &  190.07 &     0.21 &    0.03 &    +105.6 \\
 26 &   73.40 &  134.86 &  260.142 &   +48.189 &  +6.23 &   -10.48 &   73.47 &  134.81 &     0.12 &    0.09 &     -36.6 \\
 27 &   68.41 &   34.63 &  259.962 &   +45.309 &  +6.57 &   -10.10 &   68.39 &   34.51 &     0.29 &    0.13 &     -96.3 \\
 28 &  195.19 &  333.37 &  265.997 &   +53.802 &  +5.72 &   -11.28 &  195.18 &  333.44 &     0.18 &    0.07 &     +94.7 \\
 29 &  232.94 &  443.85 &  268.383 &   +56.873 &  +3.23 &   -13.47 &  232.89 &  443.75 &     0.29 &    0.11 &    -115.5 \\
 30 &  165.82 &  150.51 &  264.157 &   +48.586 &  +4.97 &   -11.89 &  165.81 &  150.62 &     0.11 &    0.11 &    +100.3 \\
 31 &  244.07 &   55.53 &  267.197 &   +45.700 &  +6.11 &   -10.51 &  244.16 &   55.46 &     0.20 &    0.12 &     -39.3 \\
 32 &  327.33 &  365.79 &  272.633 &   +54.288 &  +5.71 &   -11.26 &  327.35 &  365.72 &     0.06 &    0.08 &     -74.2 \\
 33 &  491.40 &  424.76 &  281.231 &   +54.897 &  +6.05 &   -10.87 &  491.48 &  424.57 &     0.30 &    0.21 &     -67.5 \\
 34 &  380.38 &  194.53 &  273.734 &   +49.159 &  +6.45 &   -10.33 &  380.49 &  194.43 &     0.22 &    0.15 &     -42.7 \\
 35 &   71.79 &   66.96 &  260.088 &   +46.241 &  +4.78 &   -11.94 &   71.72 &   66.98 &     0.14 &    0.07 &    +161.1 \\
 36 &  126.96 &  278.26 &  262.608 &   +52.301 &  +2.38 &   -14.11 &  127.01 &  278.21 &     0.15 &    0.08 &     -41.6 \\
 37 &  163.55 &  309.90 &  264.399 &   +53.172 &  +9.01 &    -7.92 &  163.53 &  309.75 &     0.27 &    0.15 &     -96.3 \\
 38 &  280.97 &  220.65 &  269.504 &   +50.324 &  +8.21 &    -8.69 &  280.87 &  220.56 &     0.29 &    0.14 &    -140.6 \\
 39 &  449.31 &  388.20 &  278.748 &   +54.209 &  +7.64 &    -9.17 &  449.28 &  388.35 &     0.23 &    0.15 &    +103.5 \\
 40 &  164.16 &   82.80 &  263.945 &   +46.642 &  +8.17 &    -8.75 &  163.97 &   82.90 &     0.37 &    0.21 &    +151.7 \\
 41 &  323.33 &  411.60 &  272.809 &   +55.611 &  +9.45 &    -7.31 &  323.34 &  411.65 &     0.05 &    0.05 &     +74.8 \\
 42 &   35.30 &   30.66 &  258.624 &   +45.187 &  +6.30 &   -10.01 &   35.35 &   30.71 &     0.12 &    0.08 &     +49.1 \\
 43 &   15.62 &  488.13 &  256.793 &   +58.269 &  +6.90 &    -9.83 &   15.60 &  488.13 &     0.26 &    0.03 &    +177.5 \\
 44 &   13.33 &  400.72 &  256.952 &   +55.768 &  +6.53 &   -10.17 &   13.33 &  400.68 &     0.01 &    0.04 &     -85.5 \\
 45 &  416.91 &  103.96 &  274.575 &   +46.407 &  +6.74 &    -9.89 &  416.92 &  103.94 &     0.10 &    0.02 &     -70.7 \\
 46 &  351.89 &  364.77 &  273.815 &   +54.140 &  +6.56 &   -10.24 &  351.84 &  364.75 &     0.09 &    0.05 &    -158.5 \\
 47 &  494.27 &  388.18 &  280.871 &   +53.872 &  +6.15 &   -10.70 &  494.26 &  388.23 &     0.13 &    0.06 &    +104.8 \\
 48 &  468.06 &   25.16 &  276.000 &   +43.908 &  +5.72 &   -10.77 &  467.98 &   25.20 &     0.13 &    0.08 &    +155.9 \\
 49 &  491.94 &   42.70 &  277.079 &   +44.257 &  +7.12 &    -9.36 &  491.91 &   42.71 &     0.05 &    0.03 &    +162.6 \\
 50 &  400.40 &  344.59 &  275.950 &   +53.301 &  +6.33 &   -10.58 &  400.40 &  344.59 &     0.05 &    0.01 &     +45.8 \\
 51 &   81.12 &  316.66 &  260.439 &   +53.420 &  +5.11 &   -11.57 &   81.16 &  316.66 &     0.08 &    0.04 &      -3.3 \\
 52 &  312.67 &   28.33 &  269.867 &   +44.718 &  +7.79 &    -8.89 &  312.73 &   28.39 &     0.16 &    0.08 &     +50.5 \\
 53 &  168.60 &  438.58 &  264.986 &   +56.870 &  +6.99 &    -9.79 &  168.59 &  438.61 &     0.05 &    0.04 &    +101.5 \\
 54 &  468.69 &  238.67 &  277.966 &   +49.891 &  +7.51 &    -9.26 &  468.69 &  238.60 &     0.01 &    0.06 &     -94.0 \\
 55 &  260.70 &  369.63 &  269.380 &   +54.665 &  +6.80 &   -10.01 &  260.67 &  369.60 &     0.07 &    0.04 &    -135.3 \\
	\hline 
	\end{tabular}
\end{table*}

\section{Maximum Likelihood Estimation (MLE) method of computing population and mass indices}

Population and mass indices are essential for computing meteor shower fluxes and investigating their evolutionary history. It is usually assumed that the cumulative number of meteor peak magnitudes and the cumulative logarithm of the number of meteoroids above a certain mass threshold follow single power-law distributions.

\cite{brown1996perseid} define the population index $r$ as:

\begin{equation}
    r = \frac{N(M_v + 1)}{N(M_v)}
\end{equation}

\noindent where $N(M_v)$ is the number of meteors with peak magnitude $M \leq M_v$.

Assuming that $dN$ is the number of meteoroids with a range of masses between $m$ and $m + dm$, \cite{mckinley1961meteor} defines the mass index $s$ as:

\begin{equation}
    dN \propto m^{-s}dm
\end{equation}

The population and the mass index are the slopes of the best fit lines on a log-log (magnitude or logarithm of mass vs. logarithm of the number of meteors) cumulative distribution of magnitudes or masses. These lines can be simply parameterized as:

\begin{equation} \label{eq:logline}
    f(x) = 10^{ax + b}
\end{equation}

\noindent where $x$ is either the magnitude (in case of the population index) or the logarithm of the mass (mass index). Note that we assume that the cumulative distribution is normalized to the $[0, 1]$ range.

Because all observations systems have some limiting sensitivity, the number of observed meteors is finite and detection efficiency drops off rapidly at magnitudes close to the limiting sensitivity. Thus there is only one part of the parameter space in which the power law assumption is valid. As a result, operationally a power law is only fit up to some limiting sensitivity $x_{min}$. A similar approach is used in other fields as well, although ways of estimating the value $x_{min}$ differ \citep{corral2019power}. \cite{pokorny2016reproducible} suggested fitting two parameters $x_{min}, x_{max}$ which describe the range of magnitudes, radar echo amplitudes (or equivalently masses) for which the linear approximation is valid. They argue that the addition of the upper boundary $x_{max}$ is necessary as small number statistics may skew the power law at larger amplitudes/masses. They used MultiNest \citep{feroz2009multinest} to fit their model, an advanced Bayesian regression method which can also perform robust error estimation.

In many studies \citep[e.g.][]{Brown2002b}, the meteoroid mass or population index is found by constructing a cumulative histogram and performing linear regression on binned data. This approach has many serious problems which are detailed in \cite{clauset2009power}, the most important of which is that the result may change with histogram binning, especially when data sets with a small number of meteors are used. \cite{molau2014obtaining} proposed a more advanced method which is capable of estimating the population index and the meteor flux at the same time. The drawback of this method is that it requires a running estimation of the meteor limiting magnitude from observations. It also is constructed such that estimating the exact value of the population index is rather subjective, and no details of the error analysis are provided.

\subsection{Description of the new method}

To estimate the population and the mass index robustly, in a statistical manner and with a reliable error estimate, we use the maximum likelihood estimation (MLE) method to fit a gamma distribution to the observed distribution of magnitudes and/or masses. This approach was inspired by the work of \cite{clauset2009power} where they question the validity of the power law assumption in many real-world applications. 

First, we tried to fit a Gumbel distribution to the mass distribution following \cite{blaauw2016optical}, but we found that it always produced fits with smaller p-values than the gamma distribution, under the null hypothesis the data and model distributions are identical.

Our choice of the gamma distribution for our probability distribution has a theoretical basis. If we assume that meteors appear at a rate $\lambda$ per unit time, then the time between meteors follows an exponential distribution with rate $\lambda$ while the total number of meteors in a given time span $t$ follows a Poisson distribution $P(\lambda t)$. A gamma distribution $Gamma (n, \lambda)$ is the sum of independent exponentially distributed random variables, and describes the length of time needed to observe a number of meteors $n$ \citep{akkouchi2008convolution, lawrence2008univariate}. 

By replacing the time in this analogy with magnitude or mass, the theory still applies. Given a meteor rate per unit magnitude or mass (i.e. a population or mass index), a gamma distribution describes the distribution of magnitudes or masses one expects given a certain total number of meteors. In the special case where the rate is a constant in our chosen time interval, the sum can be simplified to an Erlang distribution. Because of historical observations that show that the mass index is not constant even on short time scales for certain showers \citep{koten2014three}, we use the more general gamma distribution.

We can show the above argument is correct quantitatively. Note that in the paragraph above we implicitly assume an exponential distribution, not a power law. The power law assumption for the distribution of meteoroid masses used in meteor astronomy comes from the foundational theoretical paper by \cite{dohnanyi1969collisional} who assumed that when asteroids/meteoroids collide, the resulting fragment mass distribution follows a power law \citep{clauset2009power}:

\begin{equation}
    p(x) = \frac{s - 1}{x_{min}} \left ( \frac{x}{x_{min}} \right )^{-s}
\end{equation}

\noindent where $s$ is the scaling parameter (differential mass index in our example) and $x_{min}$ is the lower bound of the power law behavior. The power law assumption continues to propagate through the analytical derivation of \cite{dohnanyi1969collisional}, resulting in a population-wide differential mass distribution index of $s = 1.834$, assuming collisional steady-state conditions. \cite{dohnanyi1969collisional} justified the use of the power law by appealing to the experimental work of \cite{gault1963spray}, who showed that the cumulative mass-size distribution of fragmented rock from craters formed by hyper-velocity impacts in the lab follows a power-law, a result confirmed in more recent experiments and simulations \citep[e.g.][]{kun1996}. 

In reality, the number of meteors naturally tapers off as the limiting system magnitude is approached. This can be modeled as a power law distribution with an exponential cutoff. Its probability density function $p(x)$ is given by \cite{clauset2009power}:

\begin{eqnarray}
    p(x) = C f(x) \\
    f(x) = x^{-s} e^{-\lambda x} \\
    C = \frac{\lambda^{1 - s}}{\Gamma(1 - s)}
\end{eqnarray}

\noindent where $\lambda$ is the decay rate, $f(x)$ the basic functional form, $C$ the normalization constant such that $\int^{\infty}_{0} C f(x) dx = 1$, and $\Gamma$ is the gamma function. If we define a shape parameter as $k = 1 - s$, it can easily be shown that the resulting probability distribution is the gamma distribution:

\begin{equation}
    p(x) = \frac{\lambda^{k} x^{k - 1} e^{-\lambda x}}{\Gamma(k)}
\end{equation}

\begin{figure}
  \includegraphics[width=\linewidth]{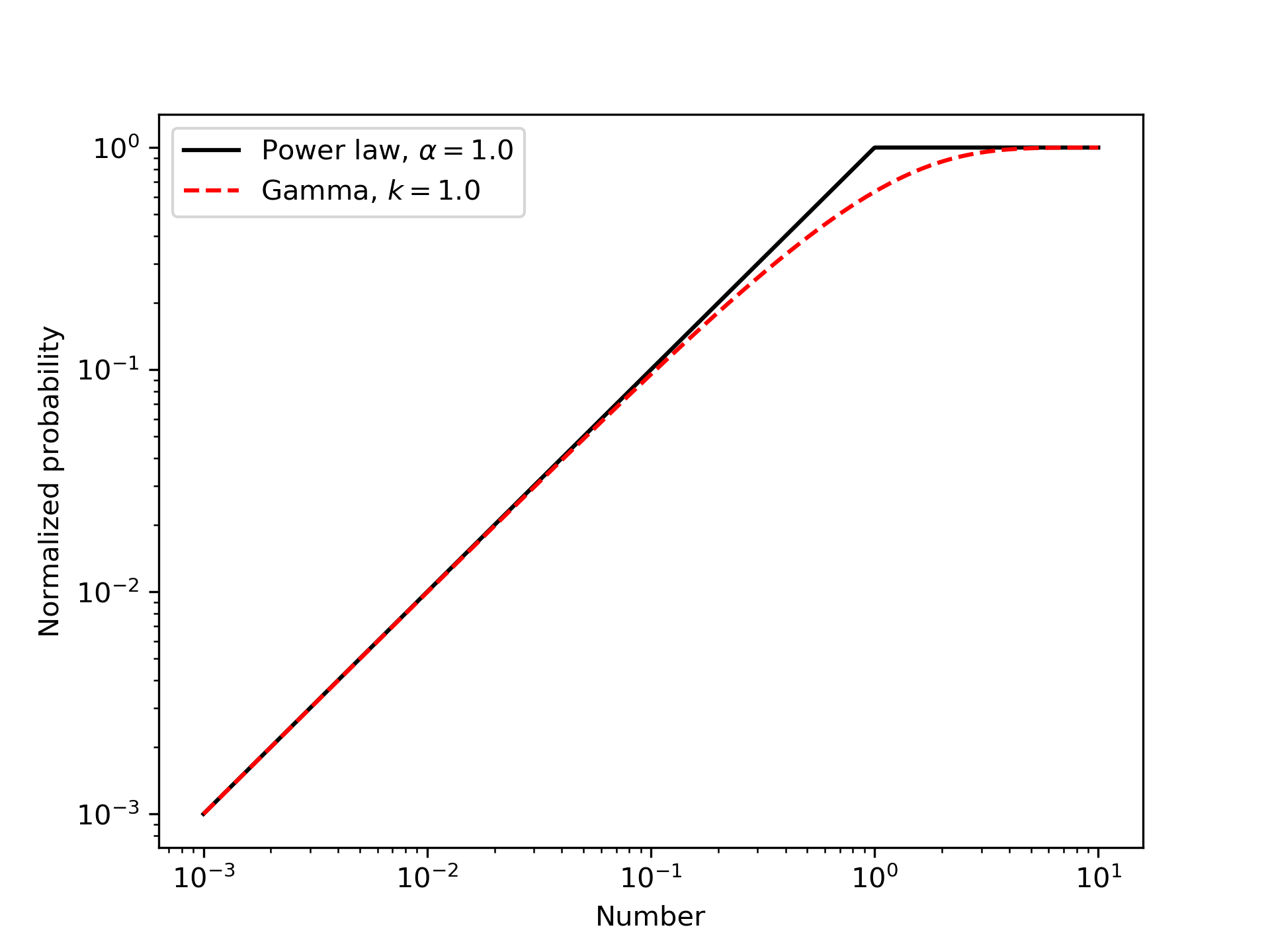}
  \caption{Comparison of the cumulative density function for a power law and a gamma function.}
  \label{fig:powerlaw_vs_gamma}
\end{figure}

Figure \ref{fig:powerlaw_vs_gamma} shows comparison of the cumulative density functions (CDF) of a power-law and of a gamma distribution. Both distributions follow the same trend at smaller values, but the exponential tapering overwhelms the power law component in the gamma distribution at around $x \sim 10^{-1}$ in this example. To estimate the power law scaling parameter from the gamma distribution, one can compute the slope of the gamma CDF before the tapering becomes significant. In theory, one may derive the slope directly as $s = 1 - k$, but in practice (especially when one works on a small data set) we find that this does not give robust estimates of population and mass indices. It appears that the fit sometimes prefers larger values of $s$ which it compensates for by using larger values of $\lambda$. For that reason, we develop a more robust method described below.

We use the gamma distribution MLE fit from the functions in the Python \texttt{scipy} library\footnote{Gamma function, \texttt{scipy} library: \url{https://docs.scipy.org/doc/scipy/reference/generated/scipy.stats.gamma.html}}.We refer readers to the documentation of the \texttt{scipy} library for details. The Python source code of our new method can be found in the Western Meteor Python Library\footnote{WMPL source code: \url{https://github.com/wmpg/WesternMeteorPyLib}} under \texttt{wmpl/Analysis/FitPopulationAndMassIndex.py}.

Figure \ref{fig:romulan_sporadic_mag_distribution} shows an example of the gamma distribution fit to meteor magnitude data. After the fit is performed, the inflection point in the probability density function (PDF) is found (the minimum of the first derivative of the PDF). This inflection point is taken as the limiting sensitivity at which data is complete; this means that the number of meteors has started decreasing at this point purely due to sensitivity losses. In our approach, the inflection point is used as an anchor; it indicates that the optimal position where the slope of the population or mass index should be estimated is at brighter magnitudes or larger masses from that point. We operationally define the optimal reference point $x_{ref}$ for:

\begin{itemize}
    \item The population index, as one magnitude brighter than the inflection point.
    \item The mass index, as 0.4 dex larger in mass than the inflection point. A 0.4 dex in mass roughly corresponds to the difference of one meteor magnitude \citep{vida2018modelling}.
\end{itemize}

\noindent Assuming an 8-bit camera is used to capture meteor data, it gives $\log_{2.512} 256 \approx 6$ magnitudes of dynamic range. As examples below show, the data is often not complete for the faintest three magnitudes, leaving only three magnitudes before reaching the point of saturation. One magnitude brighter than the completeness magnitude is a rough midpoint between the point of completeness and sensor saturation, allowing for some leeway in case the saturation point is shifted due different meteor angular speeds.

The mass index $s$ or the population index $r$ can then be computed from the slope $a$ of the log survival function $S$ (i.e. complementary cumulative distribution function) of the gamma distribution at the reference point $x_{ref}$ as:

\begin{eqnarray}
    a_s = 10^{- \frac{d}{dx} \log S_s(x_{s, ref})} \\
    s = 1 + a_s \\
    a_r = 10^{- \frac{d}{dx} \log S_r(x_{r, ref})} \\
    r = 10^{a_r}
\end{eqnarray}

\noindent where $S_s$ and $x_{s, ref}$ are fit to logarithms of mass, and $S_r$ and $x_{r, ref}$ are fit to magnitudes. Given a generic cumulative density function $F(x)$, the survival function $S(x)$ is simply:

\begin{equation}
    S(x) = 1 - F(x)
\end{equation}

\noindent and we compute the derivative in $a$ numerically. The intercept $b$ in equation \ref{eq:logline} can be computed as:

\begin{equation}
    b = \log S(x_{ref}) + a x_{ref}
\end{equation}

\noindent The Kolmogorov-Smirnov test is used to compute the p-value which we use as a measure of goodness of fit.

Next, we compute the effective meteor limiting magnitude/mass $l_m$ used in flux estimation as a point where the line $f(x) = 10^{ax + b}$ is equal to the normalized cumulative value of 1, i.e. $F(x) = 1$:

\begin{equation}
    l_m = -b/a
\end{equation}

\noindent We find that this approach is equivalent to the method of \cite{blaauw2016optical} who take the mode of the fitted Gumbel distribution as the effective limiting magnitude. We suggest that estimating the limiting meteor magnitude directly from meteor data is preferable to estimating it using stellar limiting magnitude and then applying angular velocity corrections, as our approach naturally includes all observational/instrumental biases.

Finally, the uncertainty of measured slopes is estimated using the bootstrap method \citep[page 140]{ivezic2014statistics}. The input data set is resampled with replacement, with the sample size equal to the size of the input data set, and the fit is repeated on the sample. The procedure is repeated 1000 times and population or mass indices are computed for each run. Slopes that are outside 3 sigma of the mean are rejected, and the resulting standard deviation is reported as the uncertainty.

\subsection{Testing the method}

\begin{figure}
  \includegraphics[width=\linewidth]{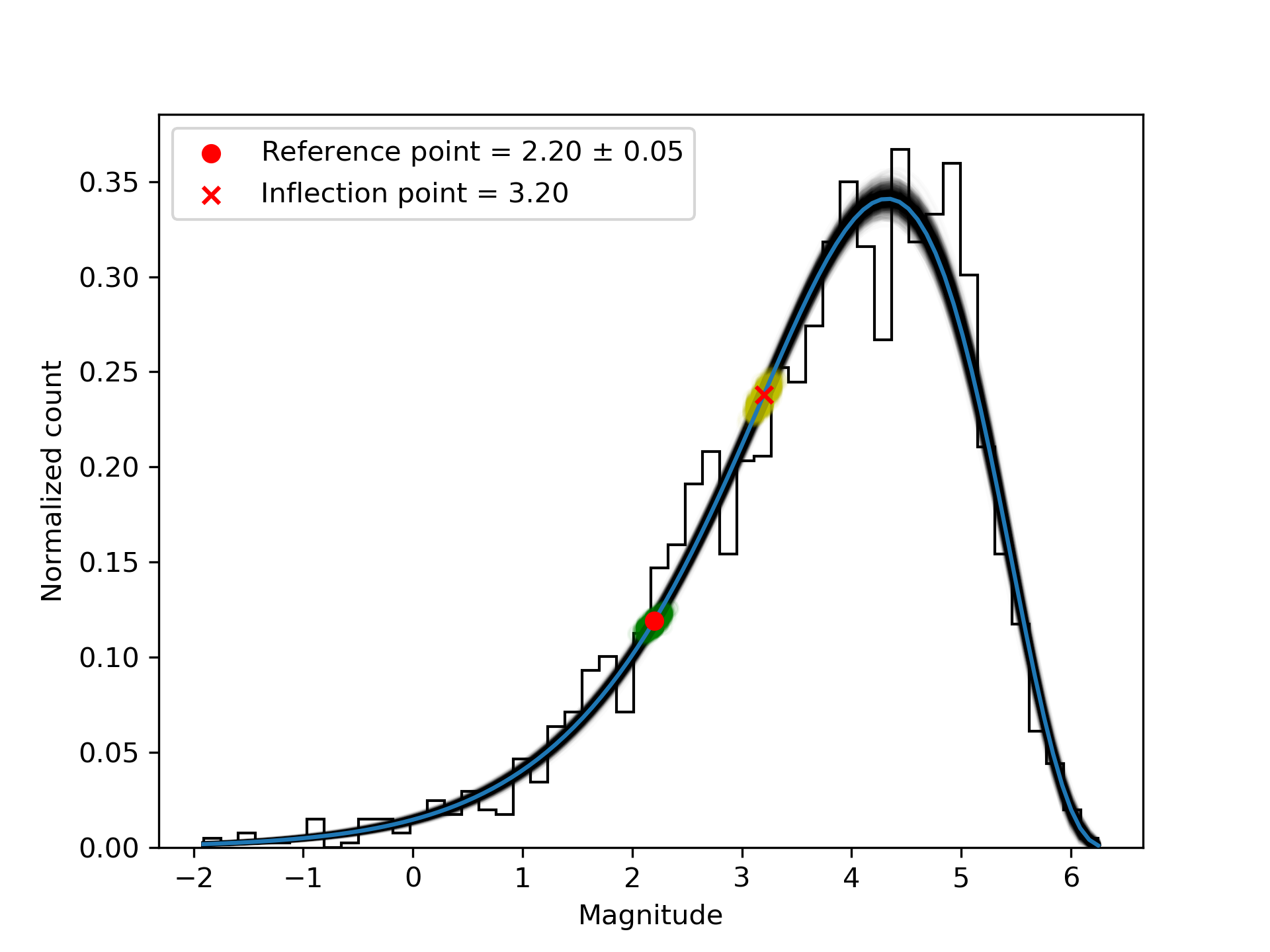}
  \caption{Distribution of peak R-magnitudes of sporadic meteors from the CAMO influx camera (histogram) and the gamma distribution fit (blue line is the main fit, black lines are all Monte Carlo runs). The jaggedness of the histogram is due to the limited precision of the magnitudes in the data file. The yellow shaded area around the inflection point represents the uncertainty in the inflection point location, and the green shaded area the uncertainty around the reference point.}
  \label{fig:romulan_sporadic_mag_distribution}
\end{figure}

Figure \ref{fig:romulan_sporadic_mag_distribution} shows the distribution of peak magnitudes of 2604 manually reduced double station sporadic meteors captured by the Canadian Automated Meteor Observatory's (CAMO) influx camera \citep{weryk2013canadian}. The data set covers a period between June 2009 and August 2014, and only those meteors which had complete light curves (both the beginning and the end was visible inside the field of view of at least one camera) were used in the analysis. We note that imposing the visibility criterion was essential, as including meteors with partially observed tracks produced a bimodal magnitude distribution which greatly skewed the results. Although the CAMO influx camera is not used for the analysis of the Draconids, we use this high quality data set used in previous work \citep[e.g.][]{campbell2015population} to demonstrate the robustness of the method. The system has a meteor limiting magnitude of $+5.5^M$, but using our approach we find that the data set is complete only up to magnitude $+3.2^M$, indicated by the inflection point, while the population index is estimated at magnitude $+2.2^M$.

\begin{figure}
  \includegraphics[width=\linewidth]{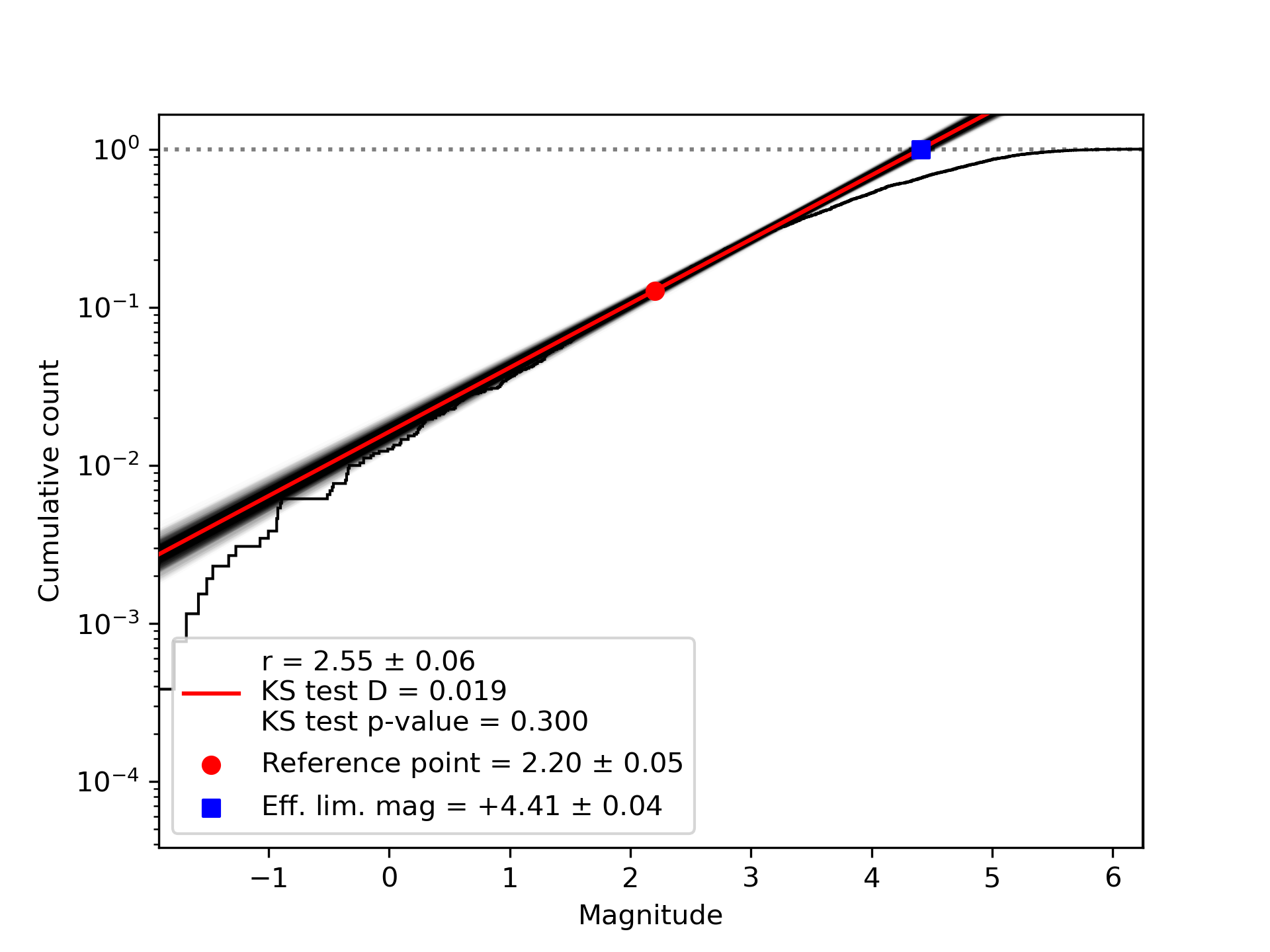}
  \caption{Cumulative distribution of peak magnitudes of sporadic meteors captured by the CAMO influx camera. The red line indicates the slope of the fitted population index.}
  \label{fig:romulan_sporadic_population_index}
\end{figure}

Figure \ref{fig:romulan_sporadic_population_index} shows the cumulative distribution of peak magnitudes and the line representing the tangent to the log cumulative distribution function of the fitted gamma function. The estimated slope of the line, i.e. the population index, is $r = 2.55 \pm 0.06$, which is consistent with other studies which have measured the expected yearly average sporadic population index from video data, e.g. \citep{molau2015population}. Note that the estimated slope follows the data histogram well. If the slope was to be estimated at the inflection point, the population index would be $r = 2.20$ and the slope would significantly deviate from the data histogram.

\begin{figure}
  \includegraphics[width=\linewidth]{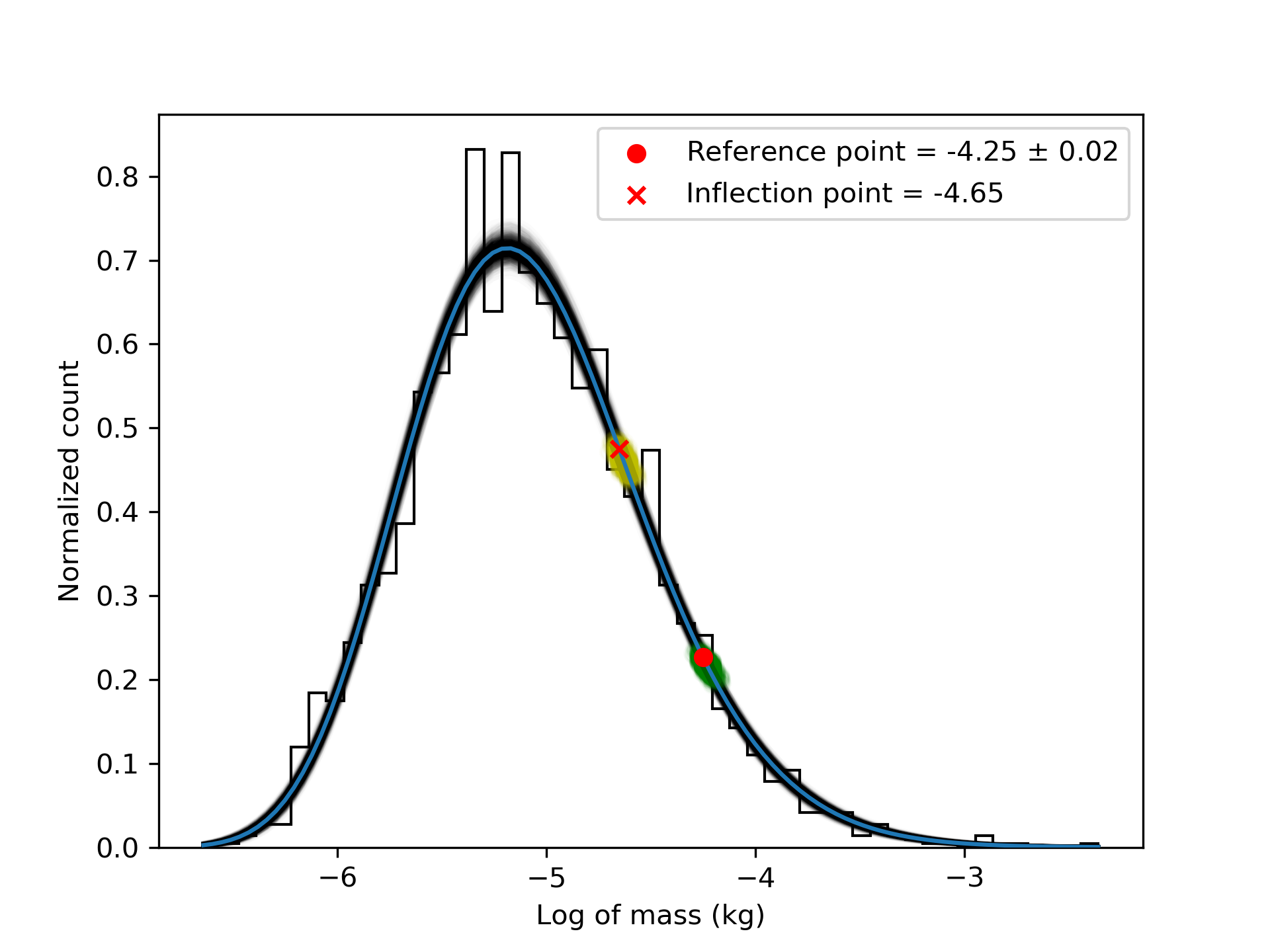}
  \caption{Distribution of masses of sporadic meteors from the CAMO influx camera (histogram) and the gamma distribution fit (blue line). Masses were computed using a fixed dimensionless luminous efficiency $\tau = 0.7\%$.}
  \label{fig:romulan_sporadic_mass_distribution}
\end{figure}

\begin{figure}
  \includegraphics[width=\linewidth]{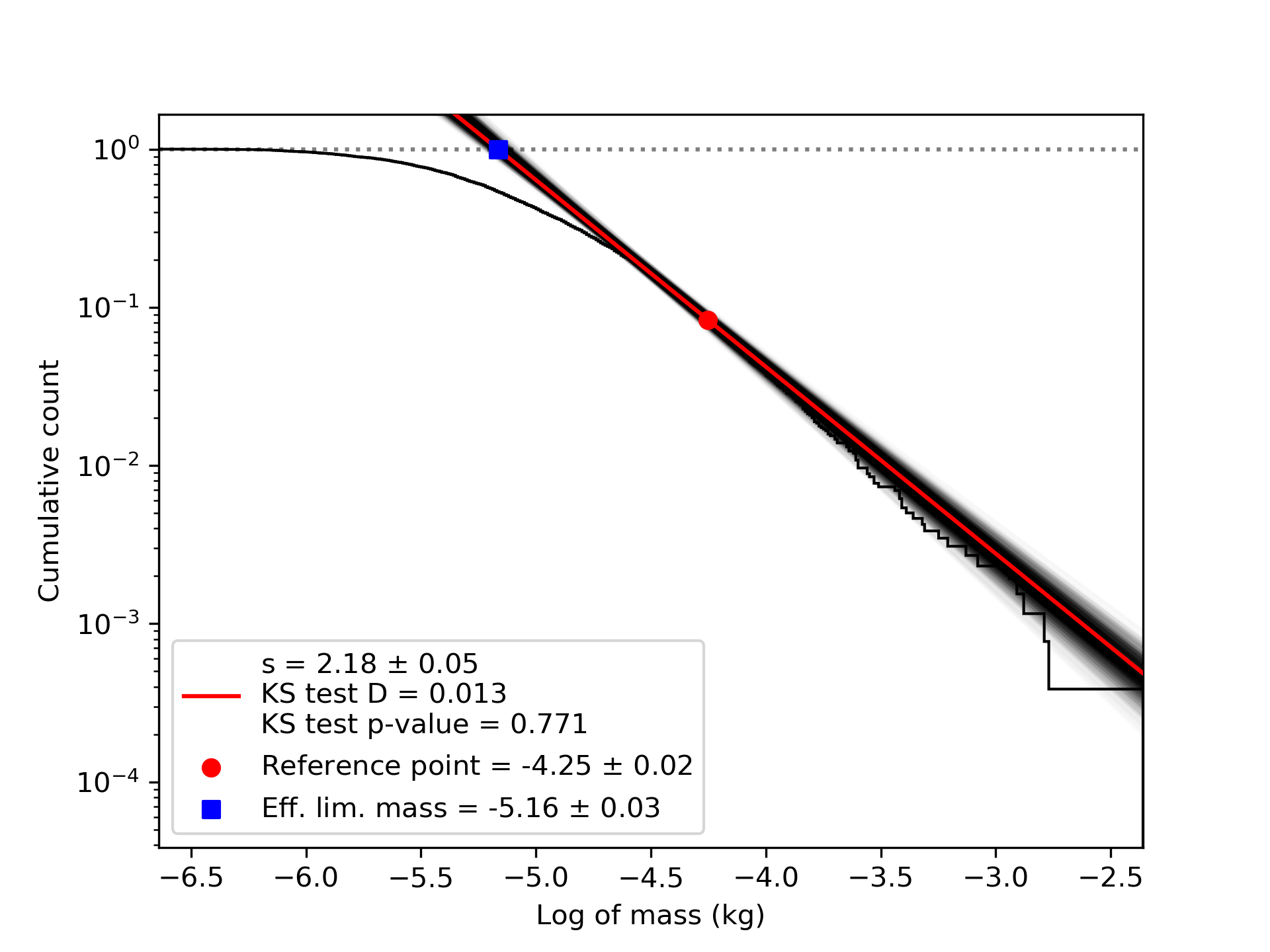}
  \caption{Cumulative distribution of masses of sporadic meteors from the CAMO influx camera. The red line indicates the slope of the fitted mass index, while the shaded black lines represent the Monte Carlo fits of the resampled set as described in the text.}
  \label{fig:romulan_sporadic_mass_index}
\end{figure}

Figures \ref{fig:romulan_sporadic_mass_distribution} and \ref{fig:romulan_sporadic_mass_index} show the distribution of masses and the cumulative histogram of masses for the same CAMO data set. The computed mass index of $s = 2.18\pm 0.05$ is very consistent with expected values from other authors \citep[e.g. ][]{blaauw2011mass}. We note that \citet{pokorny2016reproducible} used the same data set, but their algorithm fitted a line to the histogram on the part after the inflection point, slightly reducing the value of the mass index to $s = 2.09$.

\begin{figure}
  \includegraphics[width=\linewidth]{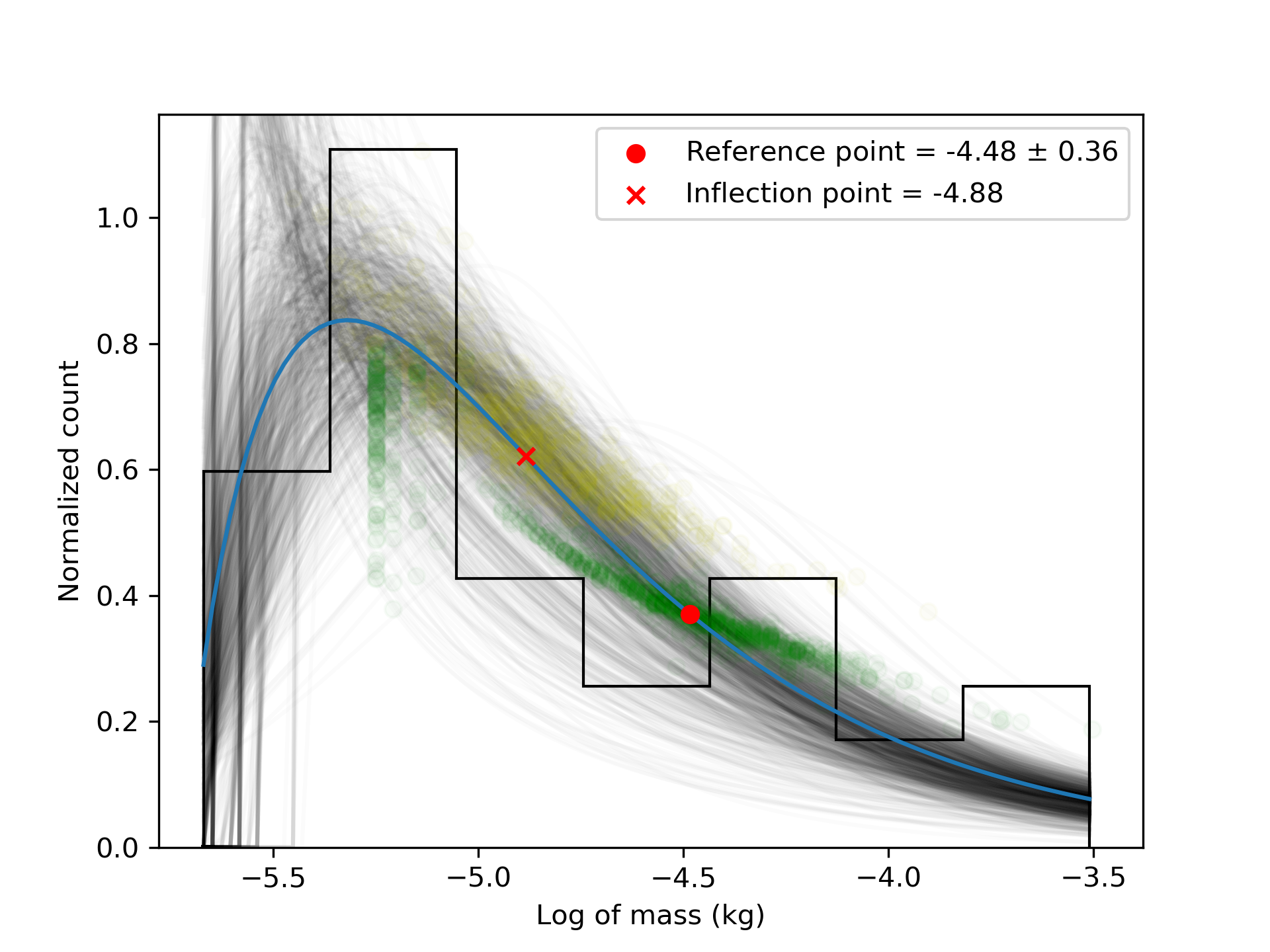}
  \caption{Distribution of masses of 38 Geminid meteors from the CAMO influx camera and the gamma distribution fit. See Fig \ref{fig:romulan_sporadic_mag_distribution} caption for the explanation of plot details.}
  \label{fig:romulan_gem_mass_distribution}
\end{figure}

\begin{figure}
  \includegraphics[width=\linewidth]{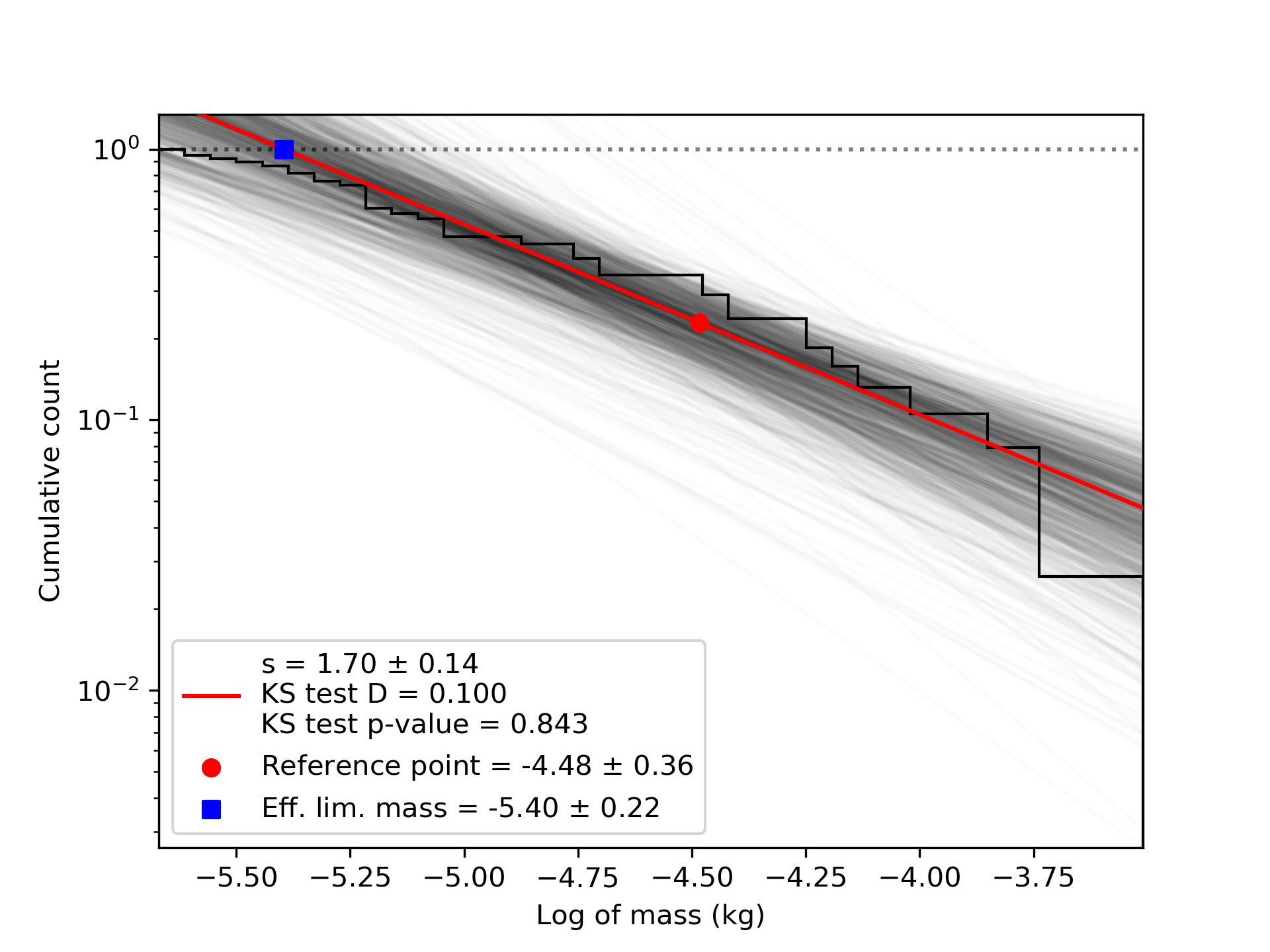}
  \caption{Cumulative distribution of masses of 38 Geminid meteors from the CAMO influx camera. The line indicates the slope of the fitted mass index. See Fig \ref{fig:romulan_sporadic_mag_distribution} caption for the explanation of plot details.}
  \label{fig:romulan_gem_mass_index}
\end{figure}

Due to the large number of data points, the estimate of the mass index of sporadic meteors can be accurately estimated, but often in practice, only tens of shower meteors are available. To demonstrate the robustness of the method, we estimate the mass index from 38 Geminids (all observed in 2012 within \ang{0.5} of solar longitude around the peak) from the CAMO influx data set (Figures \ref{fig:romulan_gem_mass_distribution} and \ref{fig:romulan_gem_mass_index}). The fit is much more uncertain, but the estimated value of the mass index for the Geminids is $s = 1.70 \pm 0.14$, which is consistent with previous work \citep{jones1982high, arlt2006activity, blaauw2011meteoroid}.

In the examples above, the p-values derived using the Kolmogorov-Smirnov test are larger than 0.05, which means that the null hypothesis that the model fits well to the data cannot be rejected with a $95 \%$ confidence level in any of the cases, even for the Geminids with a small number of meteors. This suggests that our theoretical assumptions are valid.

\begin{figure}
  \includegraphics[width=\linewidth]{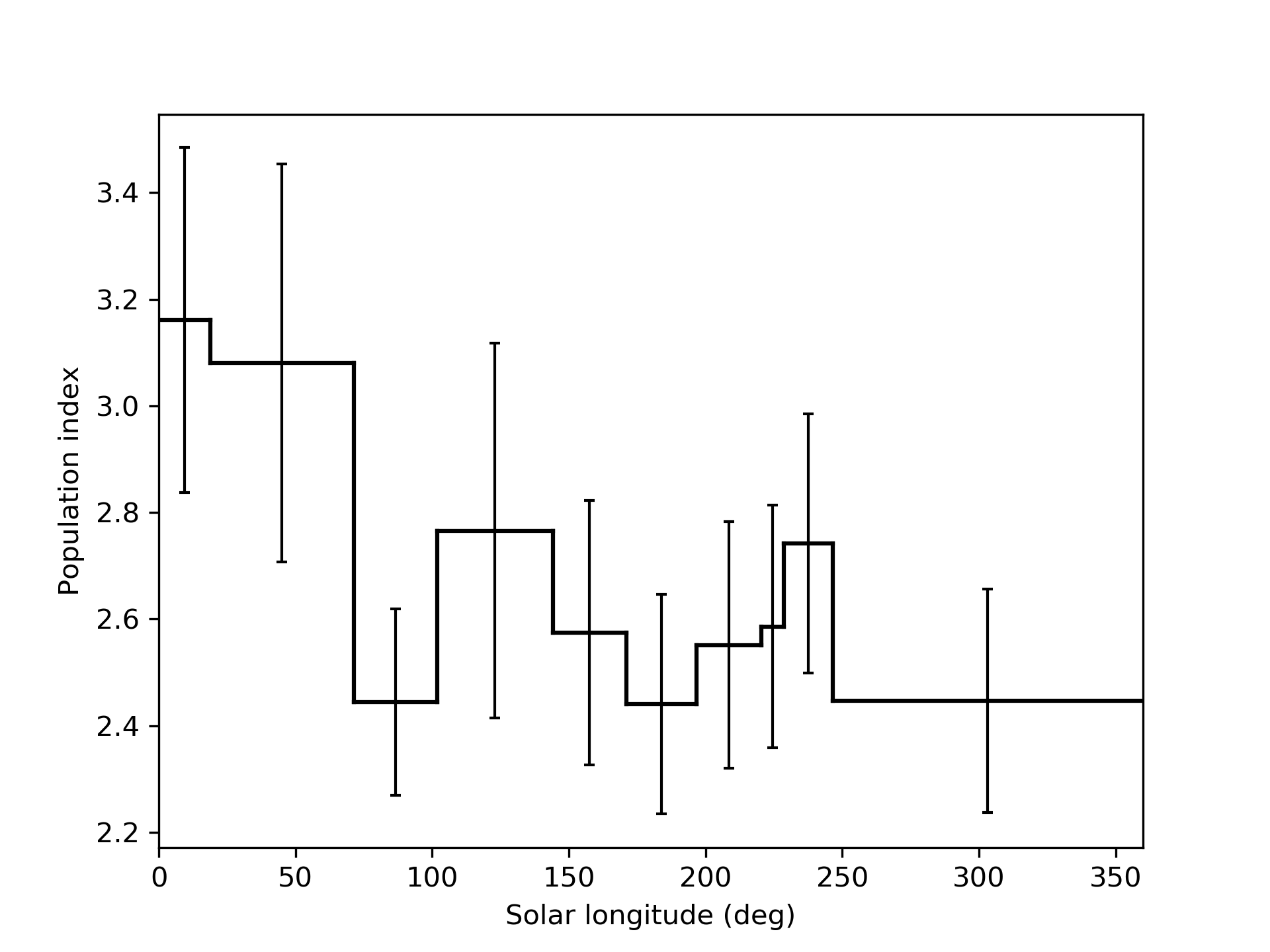}
  \caption{Annual variation in the population index for sporadic meteors using data from the CAMO influx camera.}
  \label{fig:romulan_year_population_index}
\end{figure}

\begin{figure}
  \includegraphics[width=\linewidth]{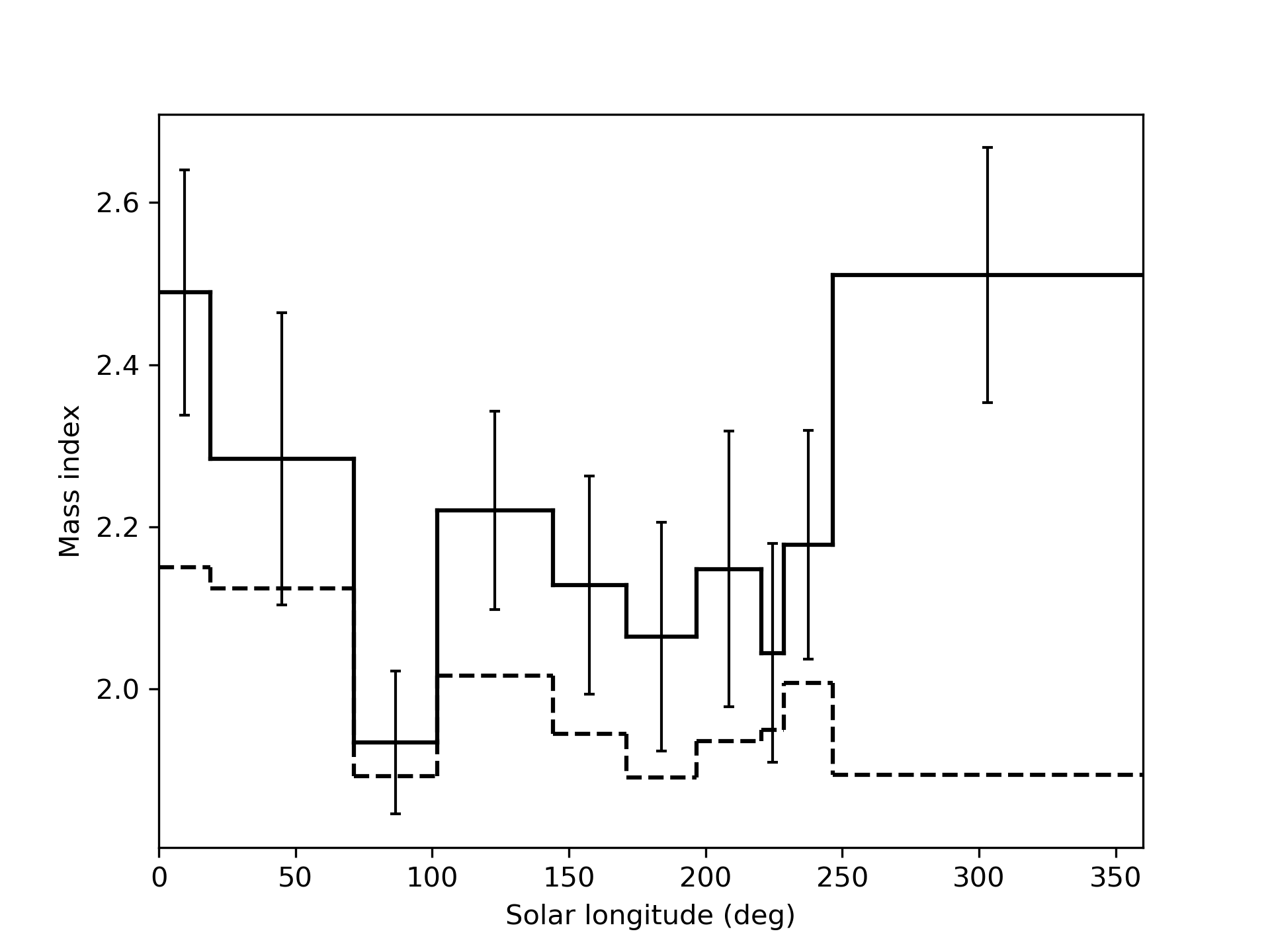}
  \caption{Annual variation in the sporadic mass index (solid line) as derived using the CAMO influx camera. The dashed line is the mass index derived from the population index using the classical relation $s = 1 + 2.3 \log r$.}
  \label{fig:romulan_year_mass_index}
\end{figure}

We investigated how the population and the mass index of sporadic meteors change throughout the year using the new method. The results are shown in Figures \ref{fig:romulan_year_population_index} and \ref{fig:romulan_year_mass_index}. The data was divided into 10 bins of equal number of meteors (around 260 in each bin) but of variable ranges of solar longitudes. Large variations in both can be seen at these optical meteoroid sizes, larger than the fairly stable values in radar sizes \citep{blaauw2011mass, pokorny2016reproducible} but following the same trend. The variations and features in the population index are comparable to those derived from visual data by \cite{rendtel2004population}, except after $\lambda_{\astrosun} > \ang{250}$ where the number of CAMO observations is very low due to bad weather conditions in winter and large bins of solar longitude had to be used. In Figure \ref{fig:romulan_year_mass_index}, the mass index estimated using our novel method is plotted as a solid line, and the mass index computed from the population index using the classical relation $s = 1 + 2.3 \log r$ \citep{mckinley1961meteor} is shown as a dashed line. Comparing the two, it can be seen that the peak magnitude alone is not a good proxy of the total meteoroid mass nor the mass distribution for sporadics.

Finally, we apply the method to EMCCD data collected on the night of August 13, 2018 ($\lambda_{\astrosun} = \ang{140}$). The meteors were detected using a hybrid cluster -  template-matching matched filter algorithm \citep{gural2007algorithms, gural2016fast}, which detects even very faint meteors close to the noise floor. The method will be described in a future paper. In total, 134 double-station Perseids were observed with the two pairs of cameras during ~6.5 hours of observation. The meteor limiting magnitude was +5.0 for these fast Perseid meteors, and only those meteors which had complete light curves  by at least one camera were used in the analysis. We measured the population index for the Perseids to be $r = 2.43 \pm 0.29$, and the mass index $s = 1.90 \pm 0.12$, comparable to long-term visual measurements at the same solar longitude \citep{rendtel2014}. In the same night, there were 283 sporadics whose population index was $r = 3.56 \pm 0.50$ and mass index $s = 2.30 \pm 0.14$, significantly higher than that of the Persieds, as expected.

\section{Mass and population index of the 2018 Draconids}

We apply the new population and mass index estimation method to the Draconids observed on Oct 8/9, 2018 using the EMCCD cameras. Meteor detection and calibration was done using RMS software \citep{vida2016open}. Instead of computing photometric masses, we computed the integrated luminous intensity and used it as a proxy for mass. 

Following \cite{vida2018modelling}, the meteoroid mass $m$ can be computed as:

\begin{eqnarray}
    I(t) = P_{0m} 10^{-0.4 M_A(t)} \\
    m = \frac{2}{v^2 \tau} \int_{0}^{t} I(t)
\end{eqnarray}

\noindent where $I$ is the luminous intensity, $P_{0M}$ is the power of a zero magnitude meteor, $M_A$ the range-corrected GAIA G-magnitude (normalized to \SI{100}{\kilo \metre}), $\tau$ the luminous efficiency, and $v$ the velocity of the meteor. $P_{0M}$ and the term before the integral simply scale the mass, and as we are only interested in the slope of cumulative logarithms of mass, the scaling has no influence on the slope if we assume that all meteors have similar velocities. This is exactly the case, as all meteors are members of the same shower. Thus, we drop these terms and compute the dimensionless integrated luminous intensity $I^*$ as a proxy for mass:

\begin{equation}
    I^* = \int_{0}^{t} 10^{-0.4 M_A(t)}
\end{equation}

The two pairs of EMCCDs recorded a total of 68 double-station Draconids with complete light curves, ranging in peak magnitudes from $+1^M$ to $+6.6^M$. We divided the data into two time bins:

\begin{enumerate}
    \item Bin 1 - Is a $\sim 30$ minute bin from 00:03:33 UTC (beginning of observation) to 00:32:01 UTC. This bin had a total of 30 Draconids and captures a part of the predicted peak.
    \item Bin 2 - Is a $\sim 60$ minute bin from 00:32:01 UTC to 01:28:58 UTC (end of observation). This bin had a total of 38 Draconids and captures the post-peak declining activity.
\end{enumerate}

\subsection{Bin 1}

\begin{figure}
  \includegraphics[width=\linewidth]{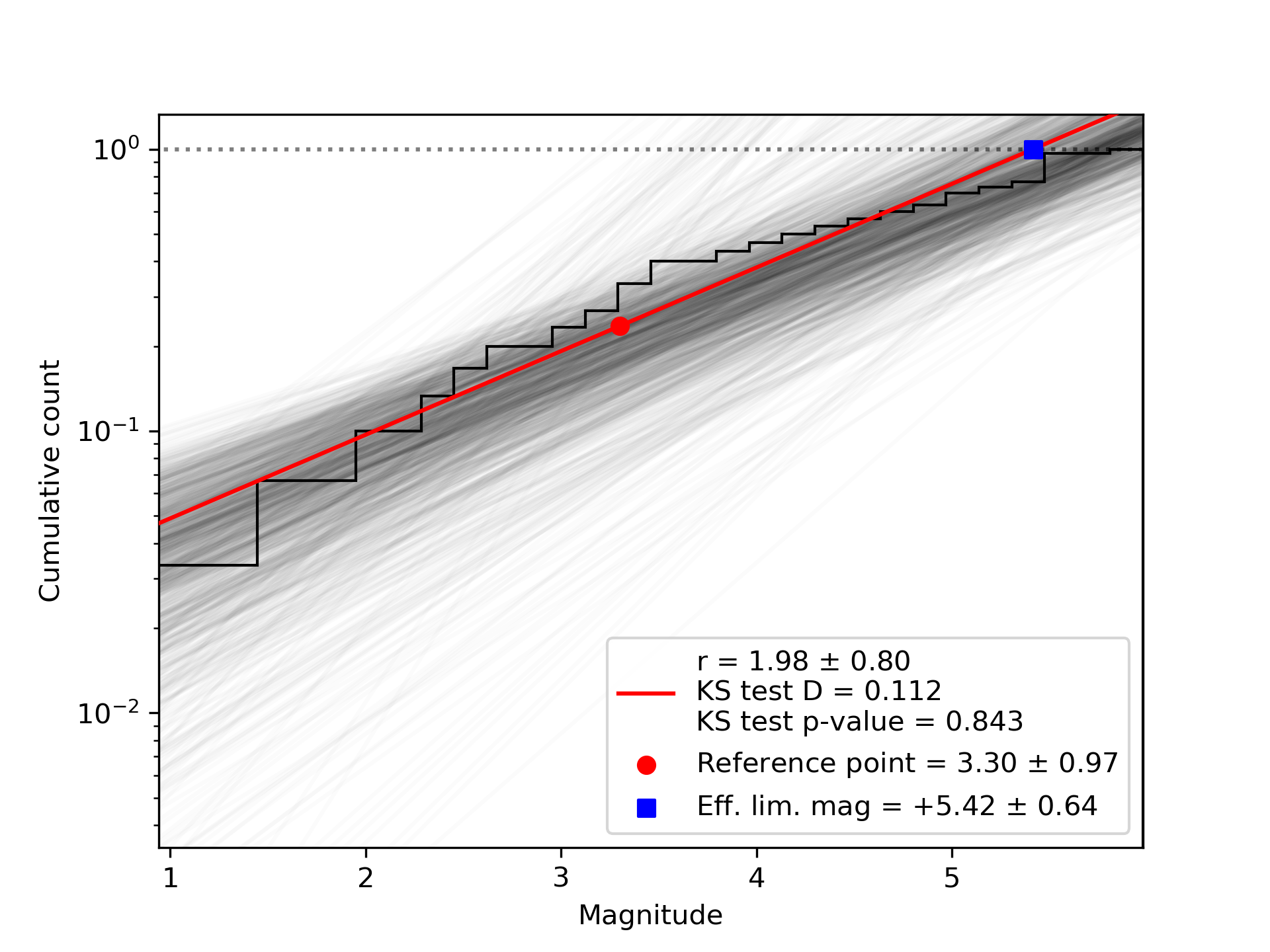}
  \caption{Cumulative histogram of magnitudes using the Draconid data in the first time bin.}
  \label{fig:emccd_2018_draconids_mag_cumulative_bin1}
\end{figure}

The first bin close to the peak, produced a population index of $1.98 \pm 0.50$ with an estimated completeness magnitude of $+4.30^M \pm 0.97$. Figure \ref{fig:emccd_2018_draconids_mag_cumulative_bin1} shows that the observations roughly follow a power-law until the reference point at $+3.30^M \pm 0.97$, after which there is a rapid drop-off in the number of observed meteors. The model fit in this case does not seem to be robust, as indicated by the uncertainty, despite the large p-value of 0.843. The population index might have been underestimated due to the overabundance of meteors with peak magnitudes of around $4^M$, possibly because of small number statistics. Alternatively, the population index might have been increasing rapidly, as evidenced by the measured value in the second bin; thus the measured value is only the average of the population index function. Because this would break the underlying assumption of a fixed power-law, this population index, the associated effective meteor limiting magnitude of $+5.42^M \pm 0.64$ and the flux derived from it are suspect.

\begin{figure}
  \includegraphics[width=\linewidth]{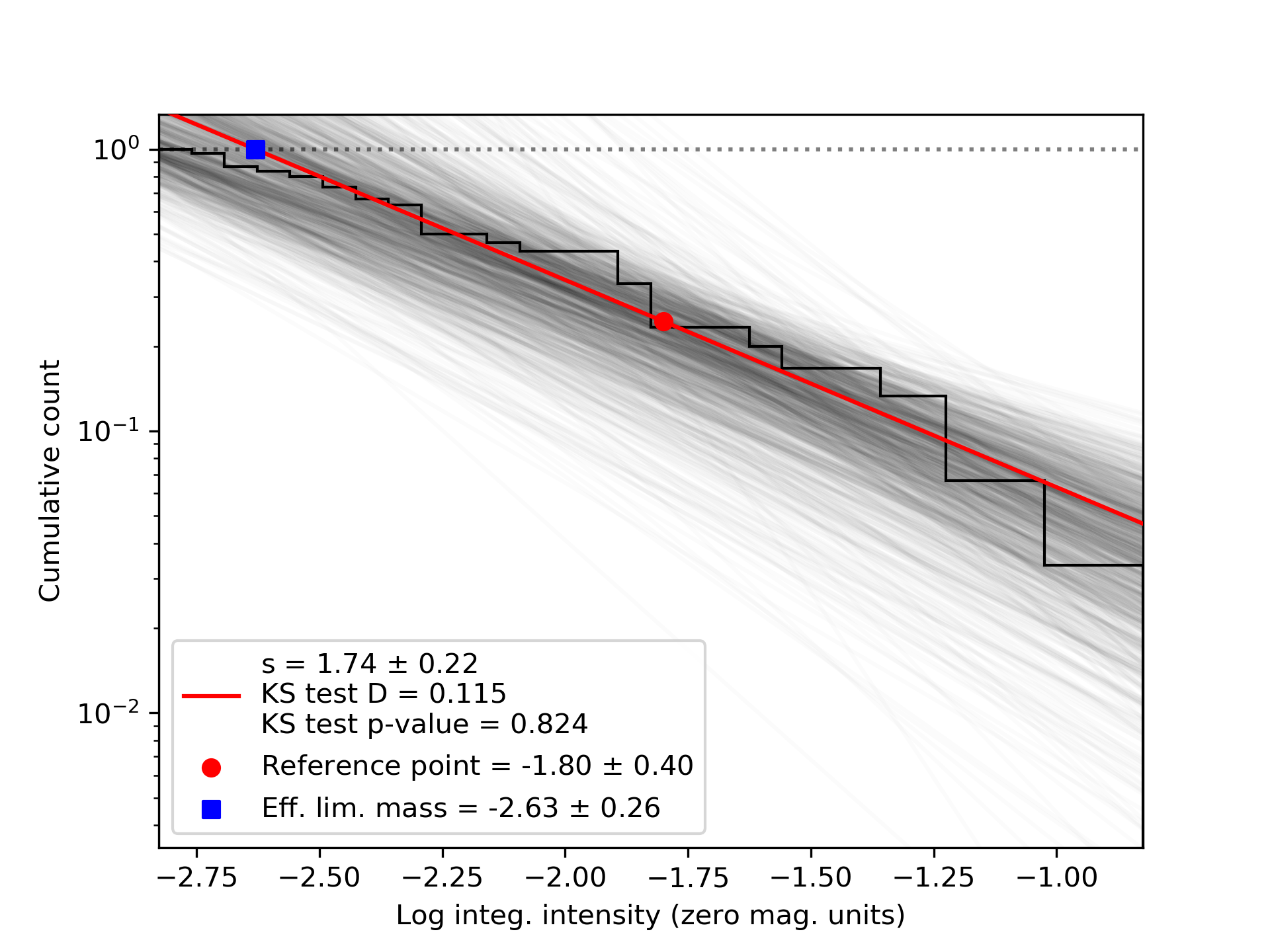}
  \caption{Cumulative histogram of integrated luminous intensities  using the Draconid data in the first time bin.}
  \label{fig:emccd_2018_draconids_mass_cumulative_bin1}
\end{figure}

Figure \ref{fig:emccd_2018_draconids_mass_cumulative_bin1} shows the cumulative histogram of dimensionless integrated luminous intensities and the mass index estimate of $1.74 \pm 0.18$ with the value of integrated intensity completeness being $-2.40 \pm 0.39$. The fit appears to be robust, and we note that the mass index estimate has a much lower uncertainty than the population index estimate. This value of the mass compares well to the value for the 2011 outburst at the peak \citep{koten2014three}.

\subsection{Bin 2}

\begin{figure}
  \includegraphics[width=\linewidth]{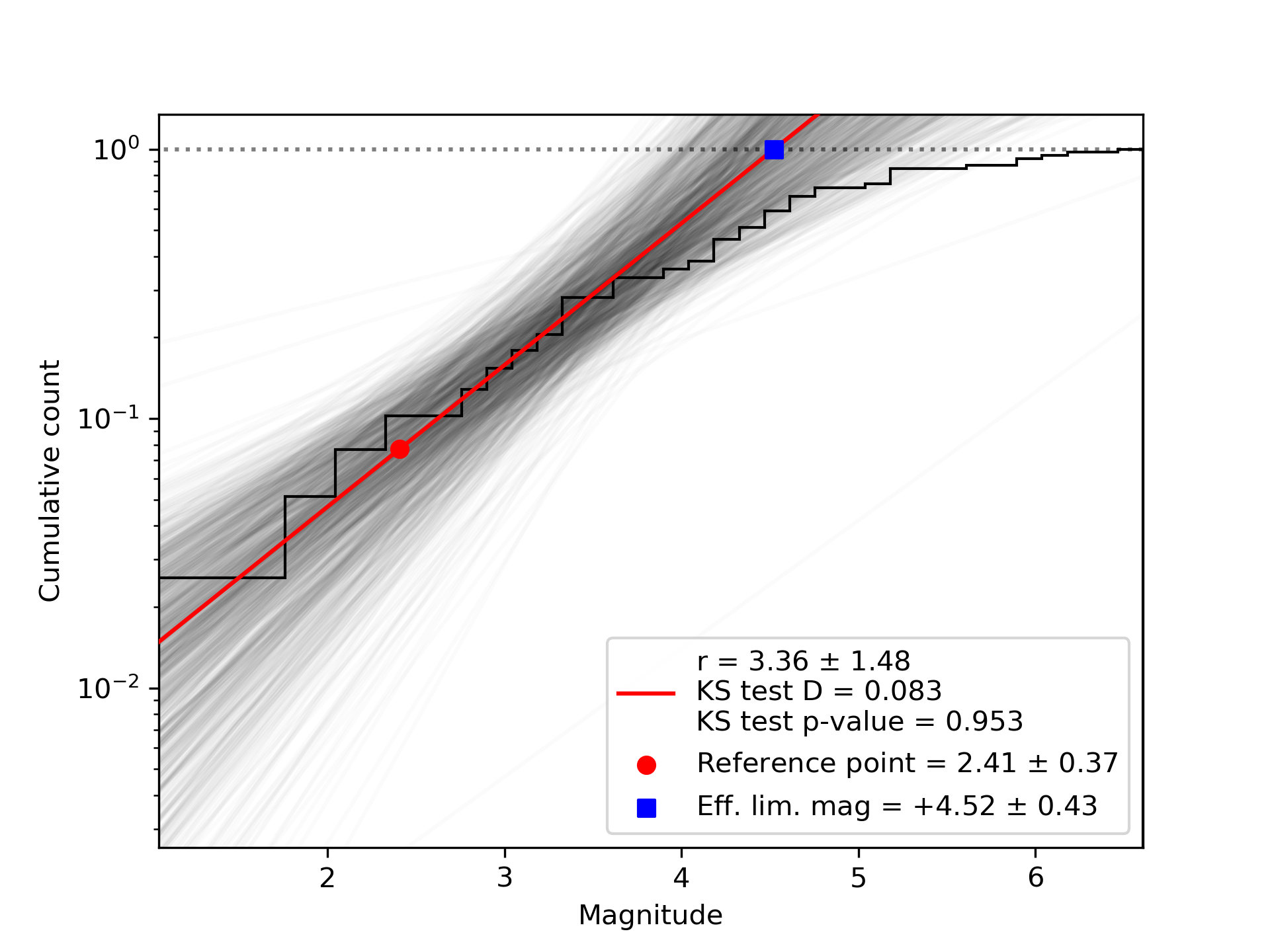}
  \caption{Cumulative histogram of magnitudes using the Draconid data in the second time bin.}
  \label{fig:emccd_2018_draconids_mag_cumulative_bin2}
\end{figure}

\begin{figure}
  \includegraphics[width=\linewidth]{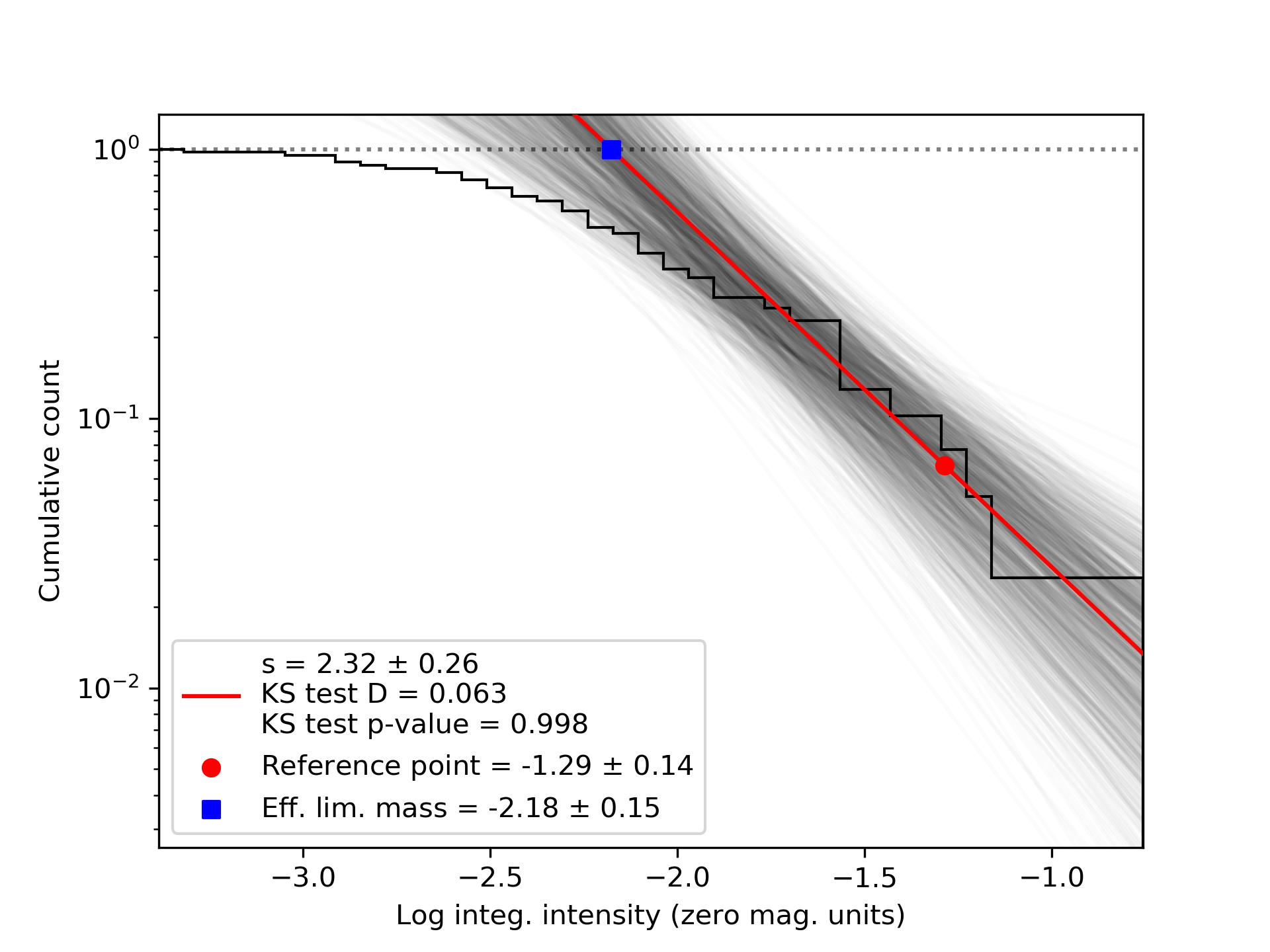}
  \caption{Cumulative histogram of integrated luminous intensities using the data in the second time bin.}
  \label{fig:emccd_2018_draconids_mass_cumulative_bin2}
\end{figure}

For the data in the second bin, we measure a much larger and more uncertain population index of $3.36 \pm 1.31$ (Figure \ref{fig:emccd_2018_draconids_mag_cumulative_bin2}), and a mass index of $2.32 \pm 0.26$ (Figure \ref{fig:emccd_2018_draconids_mass_cumulative_bin2}). The large difference from the values estimated for bin 1 indicate a fast change in the particle size distribution as the Earth moves through the stream. This rapid change is the probable cause of the increased uncertainty, and this 60 minute bin likely encompasses a range of meteoroid size distributions. We note that the speed of the mass index increase is similar to the 2011 Draconids, when the mass index changed from 1.84 to 2.30 in one hour \citep{koten2014three}.

\section{Flux} \label{sec:flux}

Collecting areas for the two pairs of EMCCD cameras were calculated using the method of \cite{campbell2016cams}. A grid of points, spaced by 4 km, was placed at heights between 92 and 98 km, to capture the typical heights of Draconid meteors. At each height, the bias for each point in the grid was calculated, taking into account the radial speed of Draconids at the time of the observation, the sensitivities of the two cameras, and the range to each camera. The area of the 4$\times$4 km region was scaled to account for meteors not seen due to lower sensitivity, using the measured mass index of the stream. An average of the areas at the heights of interest was taken, though the collecting area of both systems was relatively constant at those heights. 

The flux was calculated by dividing the number of two-station Draconids by the overlapping collecting area of each camera pair. The same two time intervals used in the mass index calculation were used for this calculation, and the flux of the first (half-hour) bin multiplied by two to get the number of meteoroids per hour. This calculation was performed for each pair of cameras separately, and then the numbers and collecting areas were added and the flux from both sets of cameras was calculated. The flux was then corrected, using the $s$ value, from the system limiting magnitudes of +5.4 and +4.5 to a limiting magnitude of +6.5, which is standard when, for example, calculating Zenithal Hourly Rates. All of these numbers, including the ZHR, are given in Table~\ref{tab:Fluxes}.

\begin{table*}[ht]
	\caption{Fluxes of Draconids, in meteoroids per square km, per hour per EMCCD camera pair (F and G). The first two rows show the measured mass index per time bin; the last two rows show the upper and lower bounds to within 1$\sigma$ of the mass index for bin 1, demonstrating the large sensitivity in flux and ZHR depending on the adopted mass index. The number of Draconids per camera pair are shown together with collecting areas per camera pair and flux.}
	\label{tab:Fluxes} 
	\centering 
	
	\begin{tabular}{lcccccccccc} 
	\hline\hline 
Bin  &   s & Num F & Num G &  F collect. & G collect. & Flux F &   Flux G &   Flux to LM &   Flux +6.5$^M$ & ZHR\\
 &  & & & area (km$^2$) & area (km$^2$) & \\
\hline
  Bin 1 & 1.74 & 14 & 13 &  328 &  687 &   0.085 &  0.038 &   0.053 &  0.11 & 586 \\
  Bin 2 & 2.32 & 22 & 29 & 214 & 420 & 0.10 & 0.071 & 0.082 & 0.99 & 681 \\
  \hline
  Bin 1 & 1.56 & 14 & 13 &  373 &  784 &   0.075 &  0.033 &   0.047 &  0.081 & 1150 \\
  Bin 1 & 1.92 & 14 & 13 &  288 &  603 &   0.097 &  0.043 &   0.061 &  0.15 & 389 \\
	\hline 
 
	\end{tabular}
\end{table*}

It is obvious that the collecting area of these systems is a strong function of the mass index: if the population index is higher (more faint meteors) the effective collecting area is smaller because many more meteors are expected to be missed in the less sensitive parts of the collecting area. We expect the flux in the second time bin to be lower than in the first, because the peak occurred just before observations began, but the flux is nearly an order of magnitude larger, and the Zenithal Hourly Rate slightly higher. We note that both camera pairs were pointed close to the apparent radiant, which may skew flux calculations. 

To get a sense of the uncertainty in the flux which comes from the uncertainty in the mass index, for the first time bin we recomputed the collecting area with the mass index varying from 1.56 to 1.92, the uncertainty limits. At the system's limiting magnitude, this changed the flux from 0.047 meteoroids km$^{-2}$ hr$^{-1}$ (at the lowest mass index) to 0.061 (at the highest); extrapolating to a limiting magnitude of +6.5 produced values from 0.081 to 0.15 meteoroids km$^{-2}$ hr$^{-1}$. The uncertainty in the ZHR is significantly higher, varying from 389 to 1150 for the two different mass indices.  There is some uncertainty in the limiting magnitude as well, which will contribute further uncertainty to these values. All of the ZHR values are higher than those observed visually by the IMO\footnote{IMO web site: \url{www.imo.net/draconids-outburst-on-oct-8-9/}}, which were of order 100. Because of the rapidly changing mass index, the flux values here should be treated with caution.

\section{Radiant distribution}

Trajectories and orbits were computed with the Monte Carlo method of \cite{vida2019meteortheory}, and the initial velocities were estimated using the sliding fit proposed in the that paper using 40\percent or more points from the beginning of the meteor. We did not apply the deceleration correction proposed by \cite{vida2018modelling} as accurate velocities were not the focus of this work, and this should not influence the radiant dispersion. 

\begin{figure}
  \includegraphics[width=\linewidth]{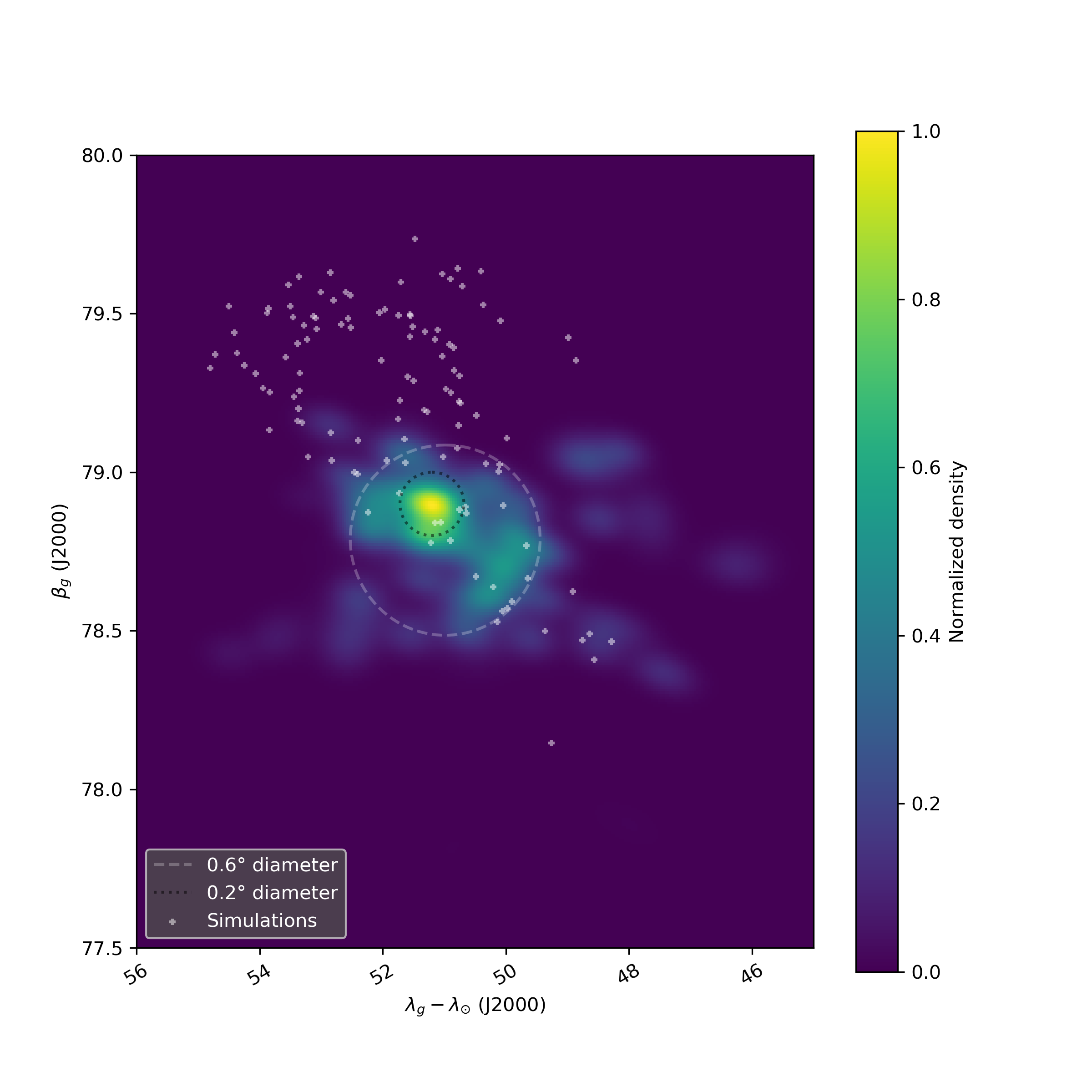}
  \caption{Sun-centered geocentric ecliptic radiant distribution of the 2018 Draconids observed with the EMCCD systems. The normalized density is color coded.}
  \label{fig:radiants_kde}
\end{figure}

\begin{figure}
  \includegraphics[width=\linewidth]{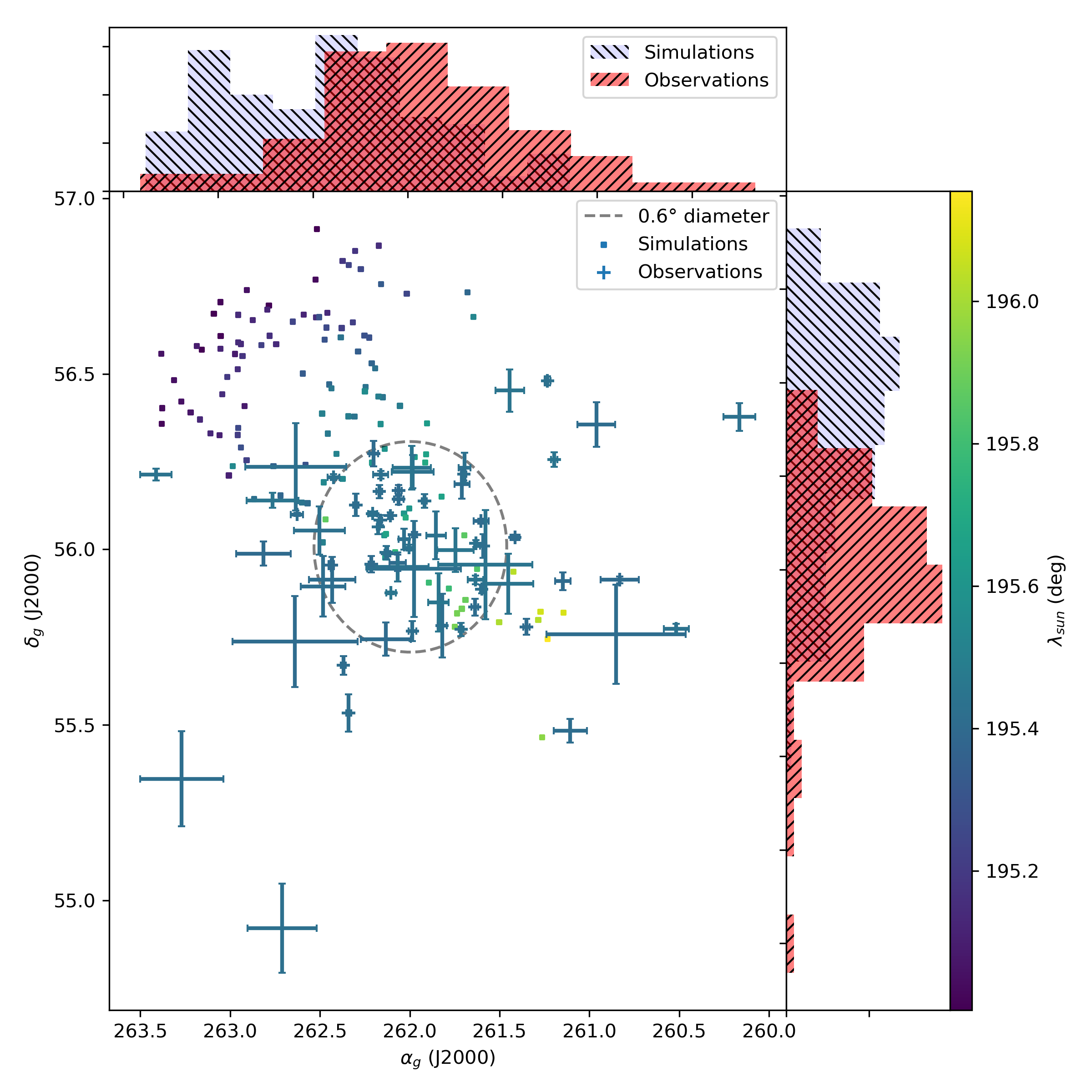}
  \caption{Geocentric radiant distribution of the 2018 Draconids observed with the EMCCD systems. The solar longitude is color coded, and the error bars show the one standard deviation uncertainty. The squares show the simulated radiants by \citep{egal2018draconid}.}
  \label{fig:radiants_sol}
\end{figure}

Figure \ref{fig:radiants_kde} shows a probabilistic radiant map in Sun-centered geocentric ecliptic coordinates generated by combining trajectory solutions of all Monte Carlo runs (200 per meteor) using kernel density estimation and a Gaussian kernel. Figure \ref{fig:radiants_sol} shows the nominal radiants with error bars in equatorial coordinates with the solar longitude color coded. Both plots include comparison between the observed and simulated radiants as produced by \cite{egal2018draconid}. In the simulation, the meteors were produced by meteoroids ejected in 1953, and we only selected simulations which hit the Earth in the range of solar longitude between $\ang{195}$ and $\ang{196}$.

The observed radiant is quite compact and the bulk of the shower falls within a circle of \ang{0.6} in diameter (dashed circle in the plot), centered around a median shower radiant $\alpha_g = \ang{261.997}$ and $\delta_g = \ang{56.007}$, or $\lambda_g - \lambda_{\astrosun} = \ang{50.985}$ and $\beta_g = \ang{78.786}$. Only radiants with high measurement uncertainties lie outside of this circle. The probabilistic radiant map shows an area of higher density at $\alpha_g = \ang{262.140}$ and $\delta_g = \ang{56.095}$, or $\lambda_g - \lambda_{\astrosun} = \ang{51.2}$ and $\beta_g = \ang{78.9}$, which is marked with a smaller circle in Figure \ref{fig:radiants_kde}, but we cannot exclude this being an artifact due to small number statistics (only 5 concentrated radiants with small uncertainty). The observed radiants match the width of simulated radiants well, but there is an offset in solar longitude. 

\begin{figure}
  \includegraphics[width=\linewidth]{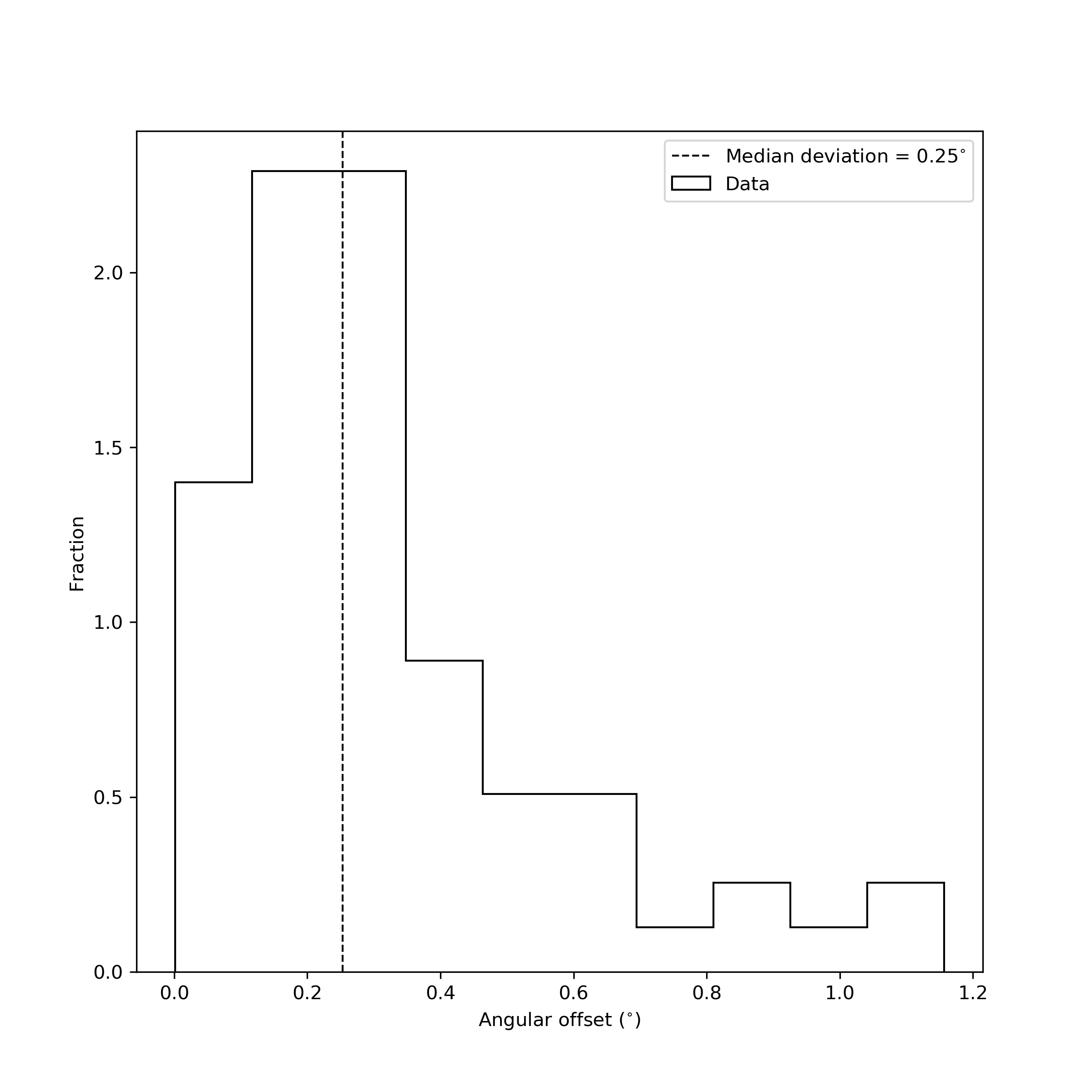}
  \caption{Distribution of angular offsets from the mean radiant of the 2018 Draconids. The vertical line marks the median angular offset of \ang{0.25}.}
  \label{fig:radiants_dists}
\end{figure}

Following \citet{kresak1970dispersion}, we compute the dispersion as the median angular offset from the mean radiant. Because the observation period was short, we did not perform the radiant drift correction. Figure \ref{fig:radiants_dists} shows a histogram of angular offsets from the mean radiant. The median angular offset was $\ang{0.25}$.

In Figure \ref{fig:radiants_sol}, the \cite{egal2018draconid} simulations show an obvious correlation of the radiant position with the solar longitude. Our observing period spanned the solar longitudes from \ang{195.400696} to \ang{195.456238}, but the simulated radiant positions matched the observations best for $\lambda_{\astrosun} \ang{\sim 195.8}$, indicating that the position of simulated particles is shifted along the orbit when compared to observations. Due to the high precision of these observations, it may possible to optimize the meteoroid ejection and integration parameters until a good match in both the position and time is achieved, improving future predictions.

\begin{figure}
  \includegraphics[width=\linewidth]{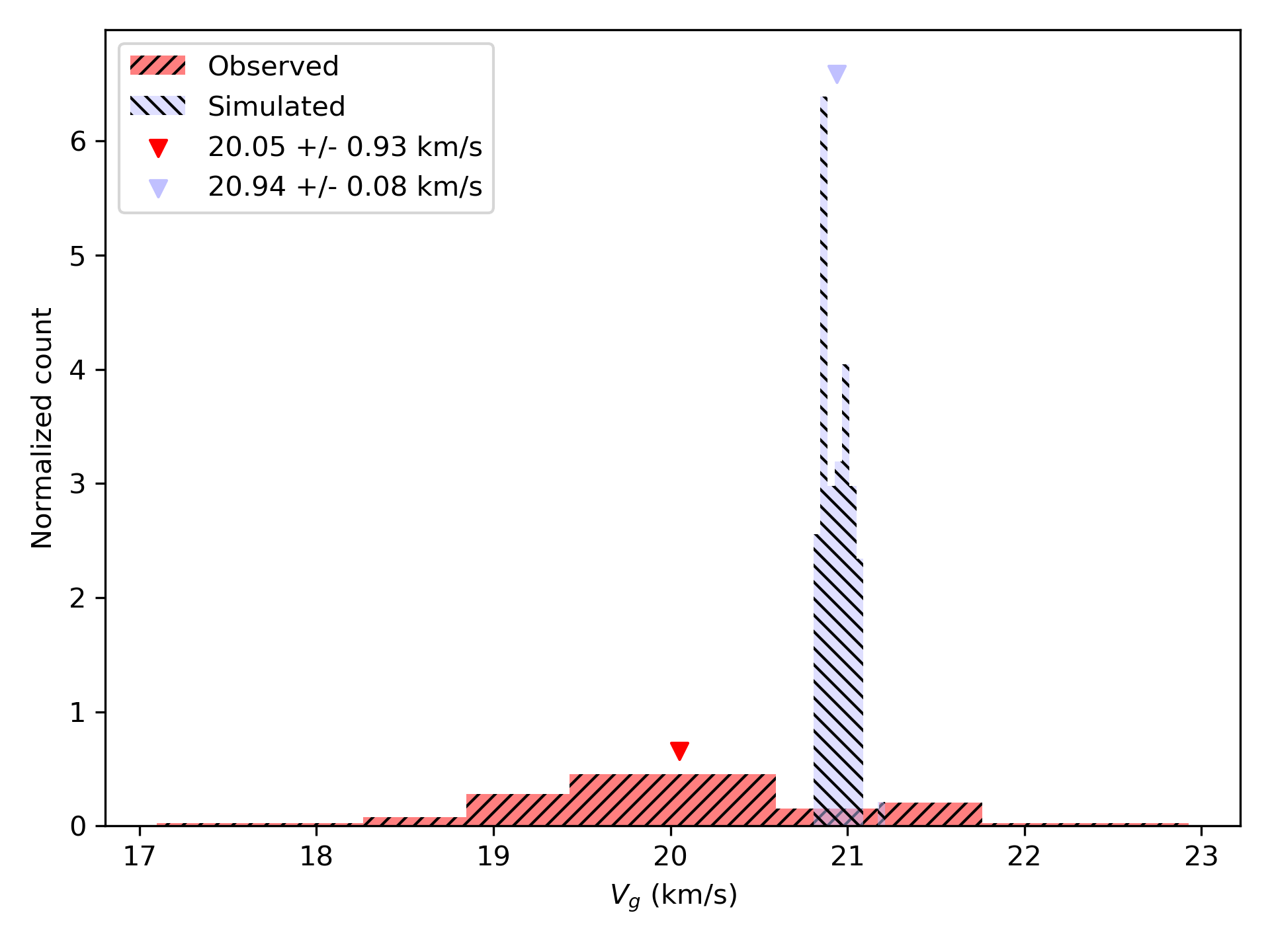}
  \caption{Comparison of simulated (blue) and observed (red) geocentric velocities.}
  \label{fig:vg_obs_sim}
\end{figure}

Figure \ref{fig:vg_obs_sim} shows histograms of simulated and measured geocentric velocities for the 2018 Draconids. The histograms were normalized so that their area is equal to 1. The mean and standard deviation were 20.05 \SI{\pm 0.93}{\kilo \metre \per \second} and 20.94 \SI{\pm 0.08}{\kilo \metre \per \second} for observed and simulated data, respectively. The observed geocentric velocities have a larger scatter than the simulations due to the limited velocity measurement precision. Furthermore, there is a systematic \SI{\sim 0.9}{\kilo \metre \per \second} underestimation of the geocentric velocity, a combined effect of meteor deceleration prior to detection \citep{vida2018modelling} and using the average velocity of the first 40\percent (or more) of the meteor as the initial velocity. In Table \ref{tab:meteor_list} we give a list of all meteors and their geocentric radiants, orbits, magnitudes, and mass proxies. We note that due to the underestimation of the geocentric velocity from lack of deceleration correction, the semi-major axes listed in the table are also underestimated.

\section{Conclusions}

We analyzed ~90 minutes of optical meteor data of the 2018 Draconid meteor shower outburst. The observations started just after the predicted peak at 00:00 UTC on October 9, 2018. We performed multi-station observations with the SOMN all-sky network and four sensitive EMCCD cameras. The all-sky system did not record any bright Draconid meteors, but 68 multi-station Draconids with complete light curves were recorded with the EMCCD systems. The astrometric and photometric calibration of the EMCCD systems has been discussed in detail, and we show that the EMCCD systems have a limiting stellar magnitude of $+10.0^M$ and an astrometric precision of ~10 arc seconds.

We developed a novel method of population and mass index computation which is based on maximum likelihood estimation. The method fits a gamma function to the distribution of magnitudes or logarithms of mass to estimate the slope of the cumulative distribution (i.e. population or mass index) before the sensitivity effects become important. We validated this approach on past data obtained with the CAMO influx system. The method also enables robust error estimation.
    
We divided EMCCD data of the 2018 Draconid outburst into two time bins of approximately equal number of meteors: the first bin spanned the time from 00:00 UTC to 00:30 UTC, and the second bin spanned from 00:30 UTC to 01:30 UTC, Oct 9, 2018. The mass index in the first bin was $1.74 \pm 0.18$, and in the second bin $2.32 \pm 0.27$, an increase of similar magnitude and rate of change to the 2011 Draconids \citep{koten2014three}.

Because of the rapid rate of change of the mass index of the shower and small number statistics, the values derived for the mass index and flux are very uncertain, particularly for the first time bin. The fluxes in the two bins were 0.11 and 0.99 meteoroids km$^{-2}$ hr$^{-1}$, respectively, corrected to a limiting meteor magnitude of +6.5. From visual observations, we expect the flux was actually falling from the first to the second time interval; the reversed trend in this data is most likely due to the assumption of a single power law in each time bin, which does not adequately describe the data. For this reason, we do not attempt to describe the flux profile of the shower. 

The measured shower radiant diameter was \ang{\sim 0.6} which matches simulations by \cite{egal2018draconid} well, but there is an offset in solar longitude of \ang{0.4}. The measured radiant dispersion is a good match to the 2011 outburst, both the high-precision radiant measurements by \cite{borovivcka2014spectral}, and the meteoroid ejection simulations by \cite{vida2019meteorresults}.

\subsection{Complete input data and trajectory solutions for EMCCD meteors}

We provide the input data and detailed trajectory solutions for the Draconids observed with the EMCCD systems, in electronic form at the CDS via anonymous ftp to \url{cdsarc.u-strasbg.fr} (\url{130.79.128.5})
or via \url{http://cdsweb.u-strasbg.fr/cgi-bin/qcat?J/A+A/}.

\begin{acknowledgements}

We thank Dr. Gunther Stober for helping with the equipment during the observing campaign. Furthermore, we thank Dario Zubovi\'c, Damir \v{S}egon, Aleksandar Merlak, Damjan Nemarnik, Vladimir Jovanovi\'{c}, Peter Eschman, Dr. Dan Kinglesmith, Przemek Naganski, Dmitrii Rychkov, Stanislav Korotkiy, and Milan Kalina for supporting the RMS project.

This work was supported in part by the NASA Meteoroid Environment Office under cooperative agreement 80NSSC18M0046. Funding from the Natural Sciences and Engineering Research Council of Canada (Grant no. RGPIN-2016-04433) and the Canada Research Chairs Program is gratefully acknowledged.

\end{acknowledgements}




\bibliographystyle{aa}
\bibliography{MAIN} 




\begin{appendix}
\section{Light curves of selected Draconids} \label{appendix:lightcurves}

\begin{figure*}
  \includegraphics[width=.5\linewidth]{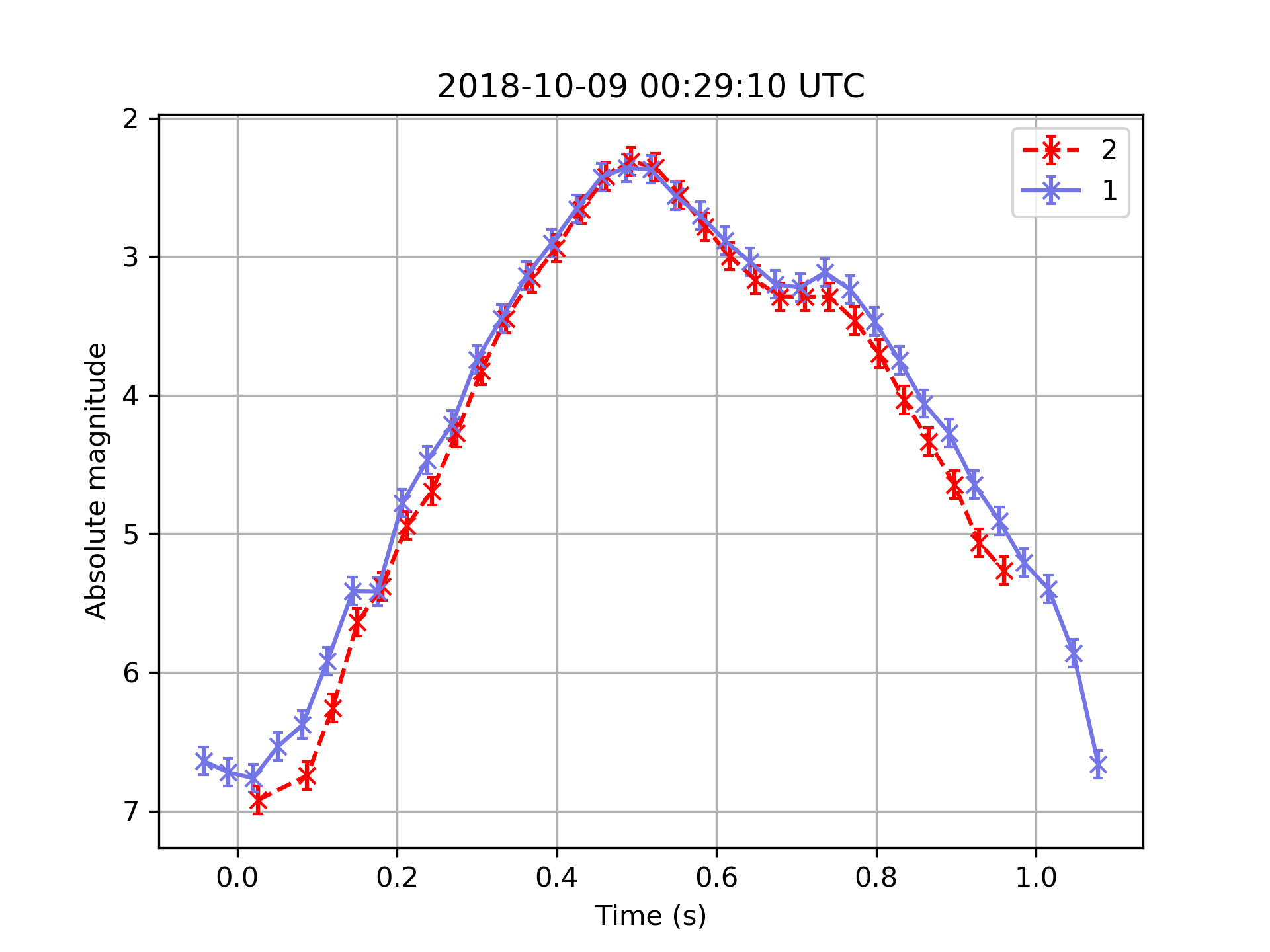}\hfill
  \includegraphics[width=.5\linewidth]{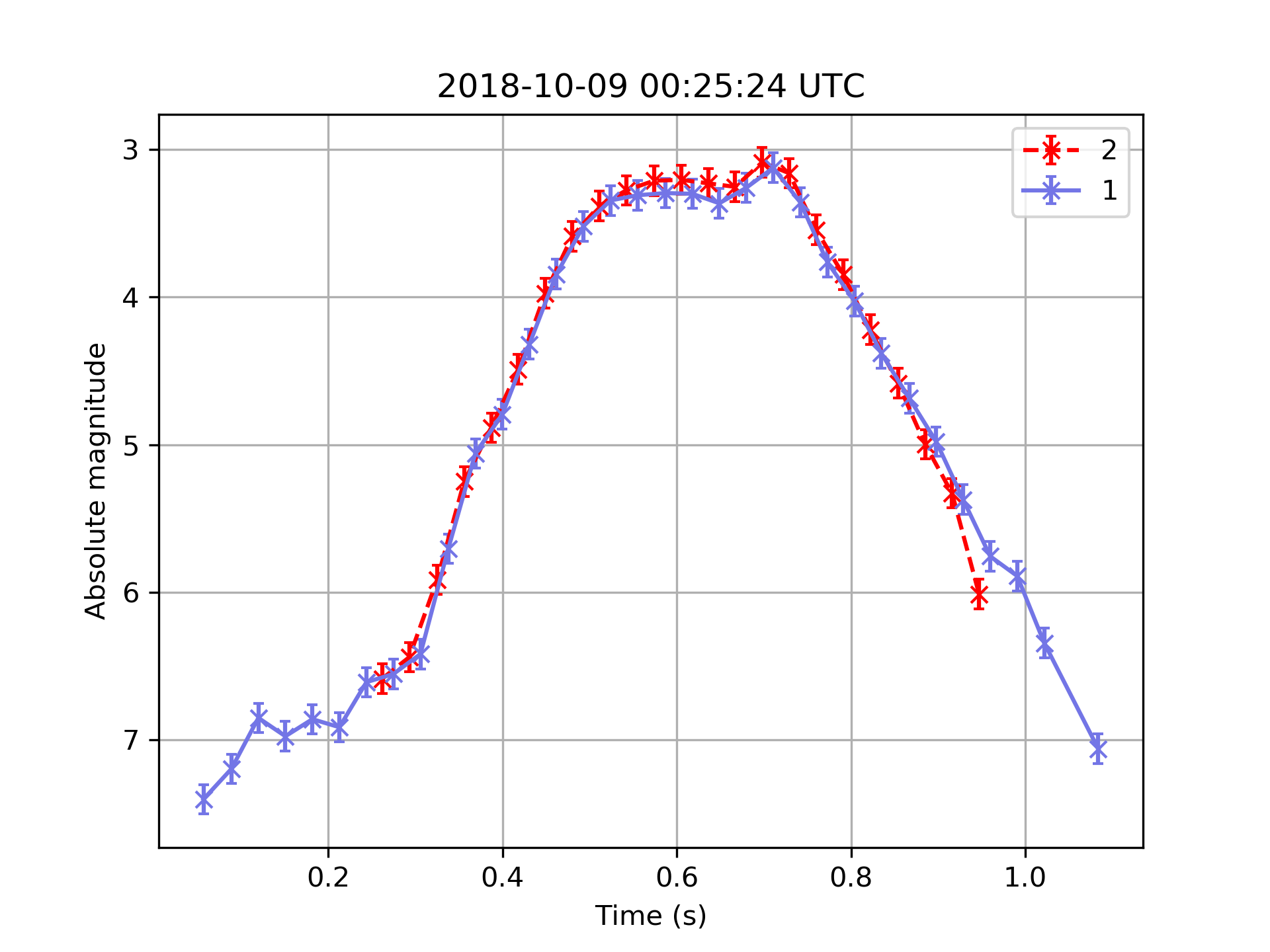}
  \includegraphics[width=.5\linewidth]{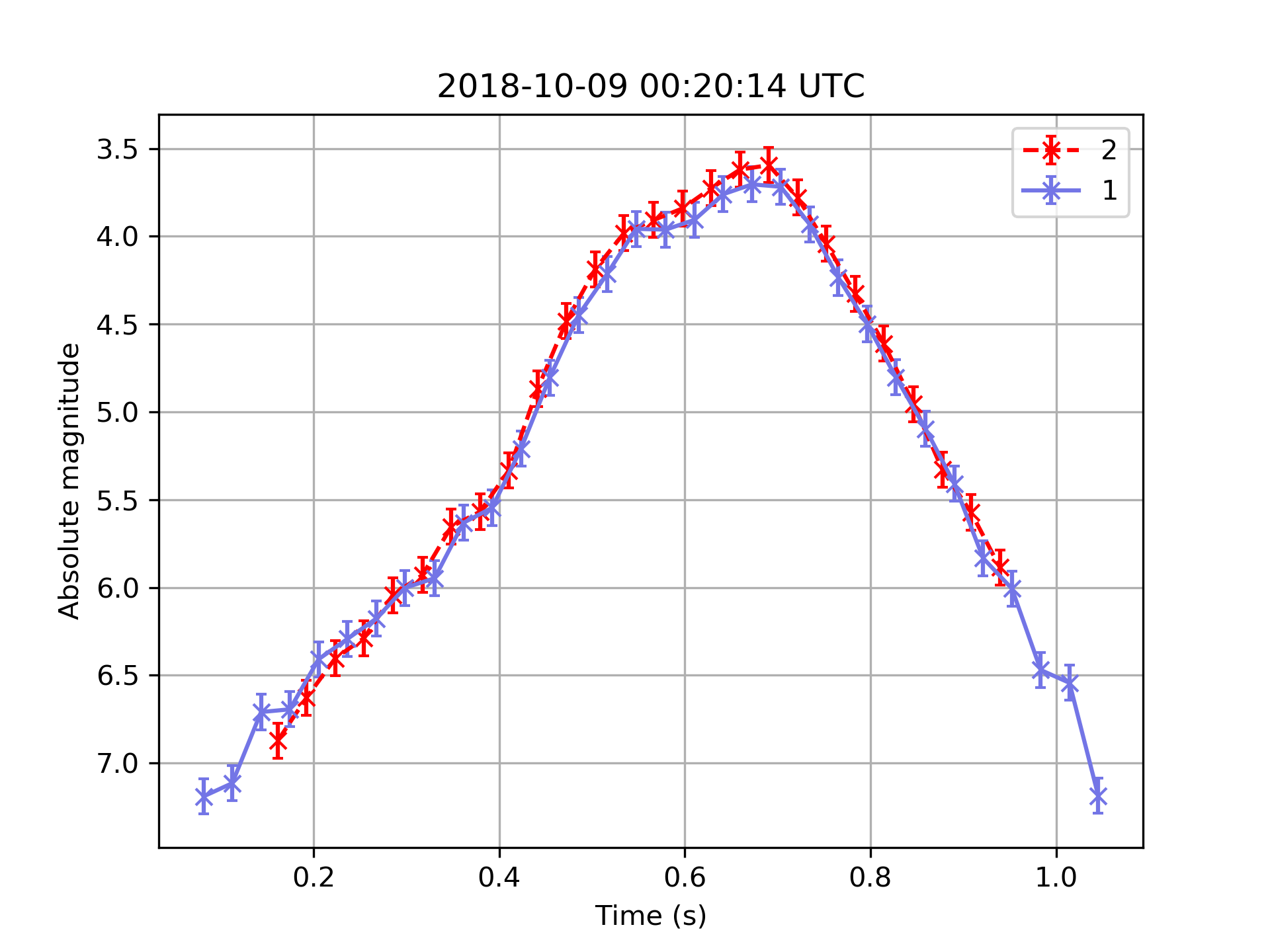}\hfill
  \includegraphics[width=.5\linewidth]{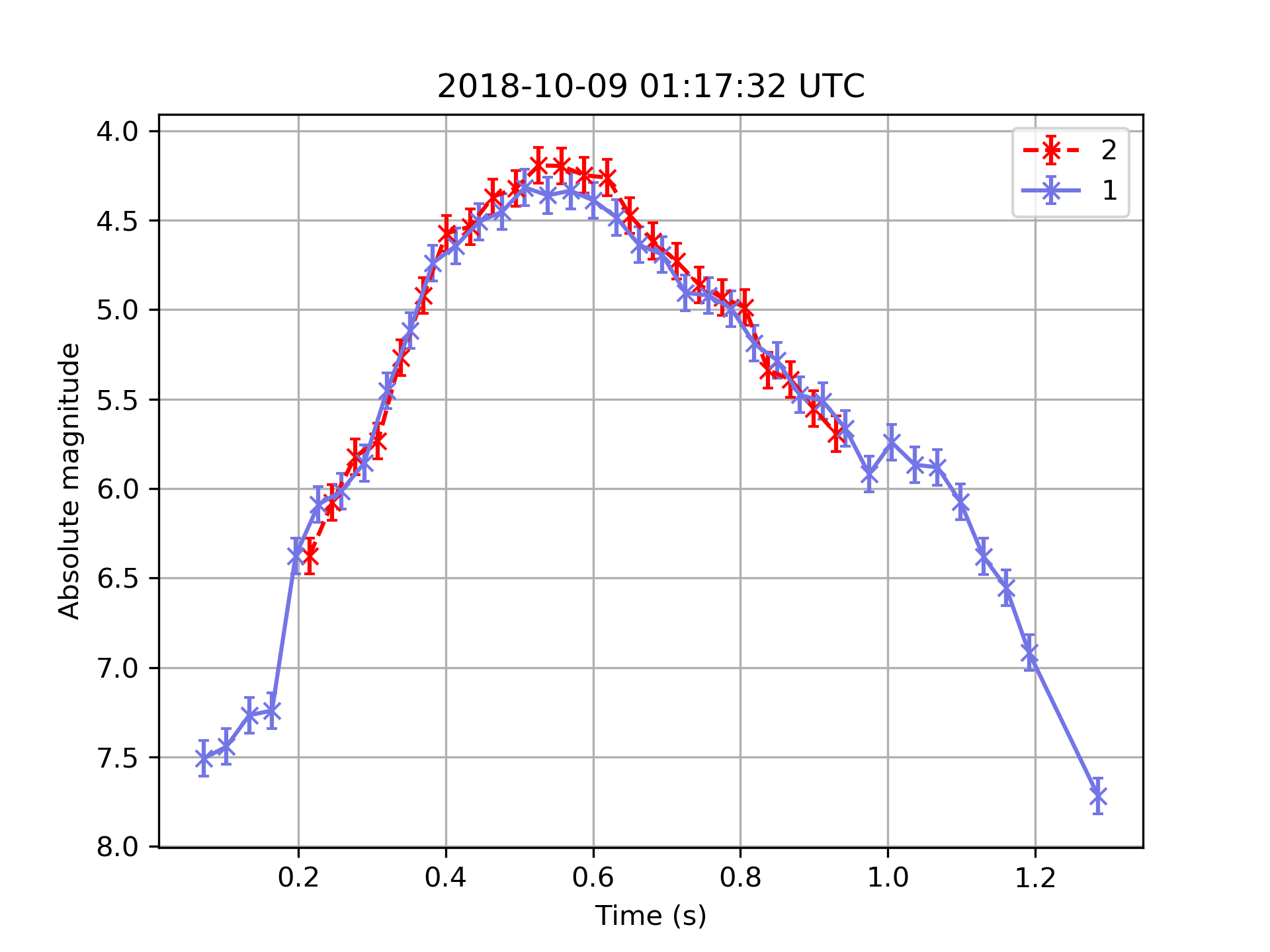}
  \includegraphics[width=.5\linewidth]{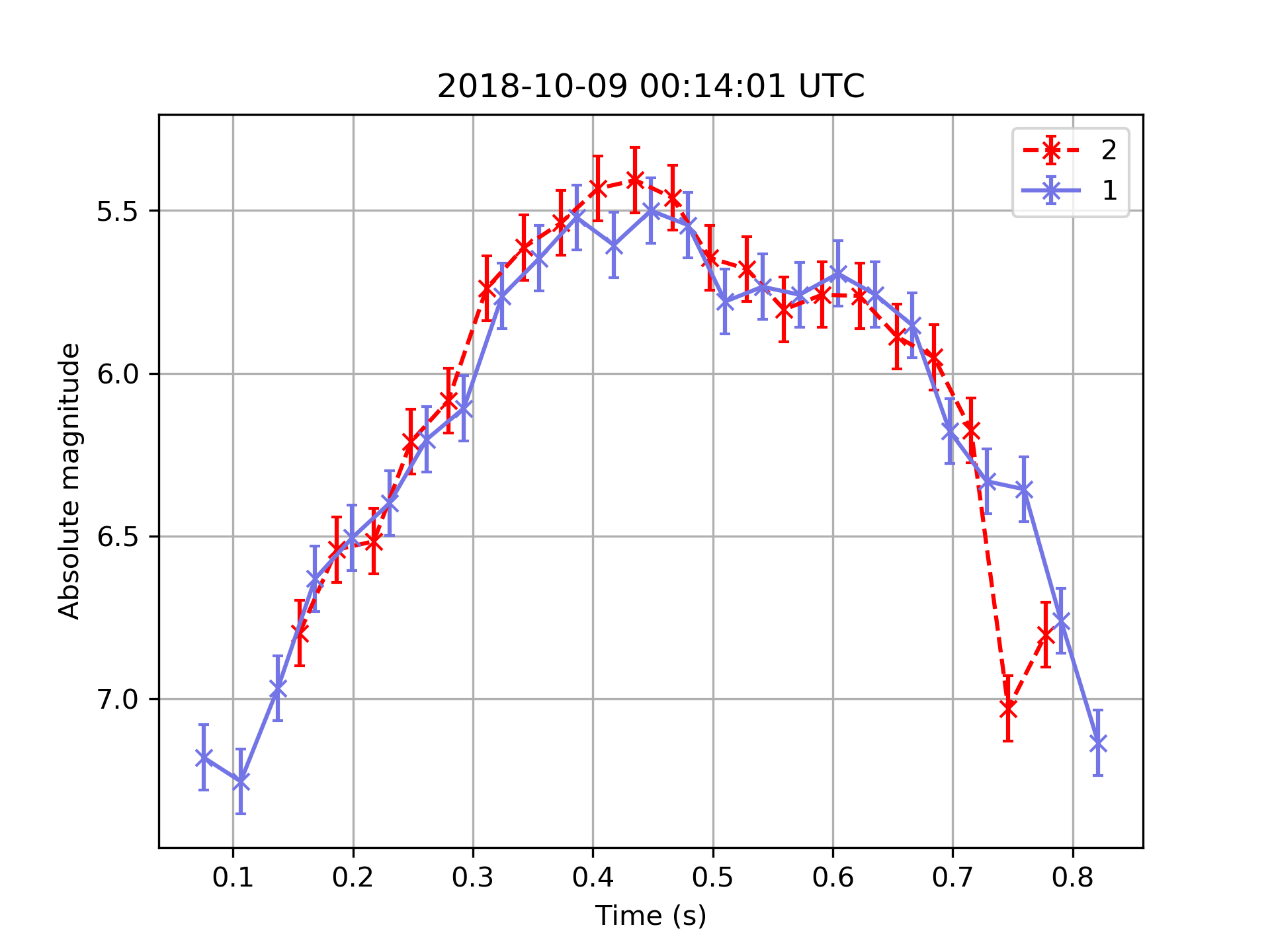}\hfill
  \includegraphics[width=.5\linewidth]{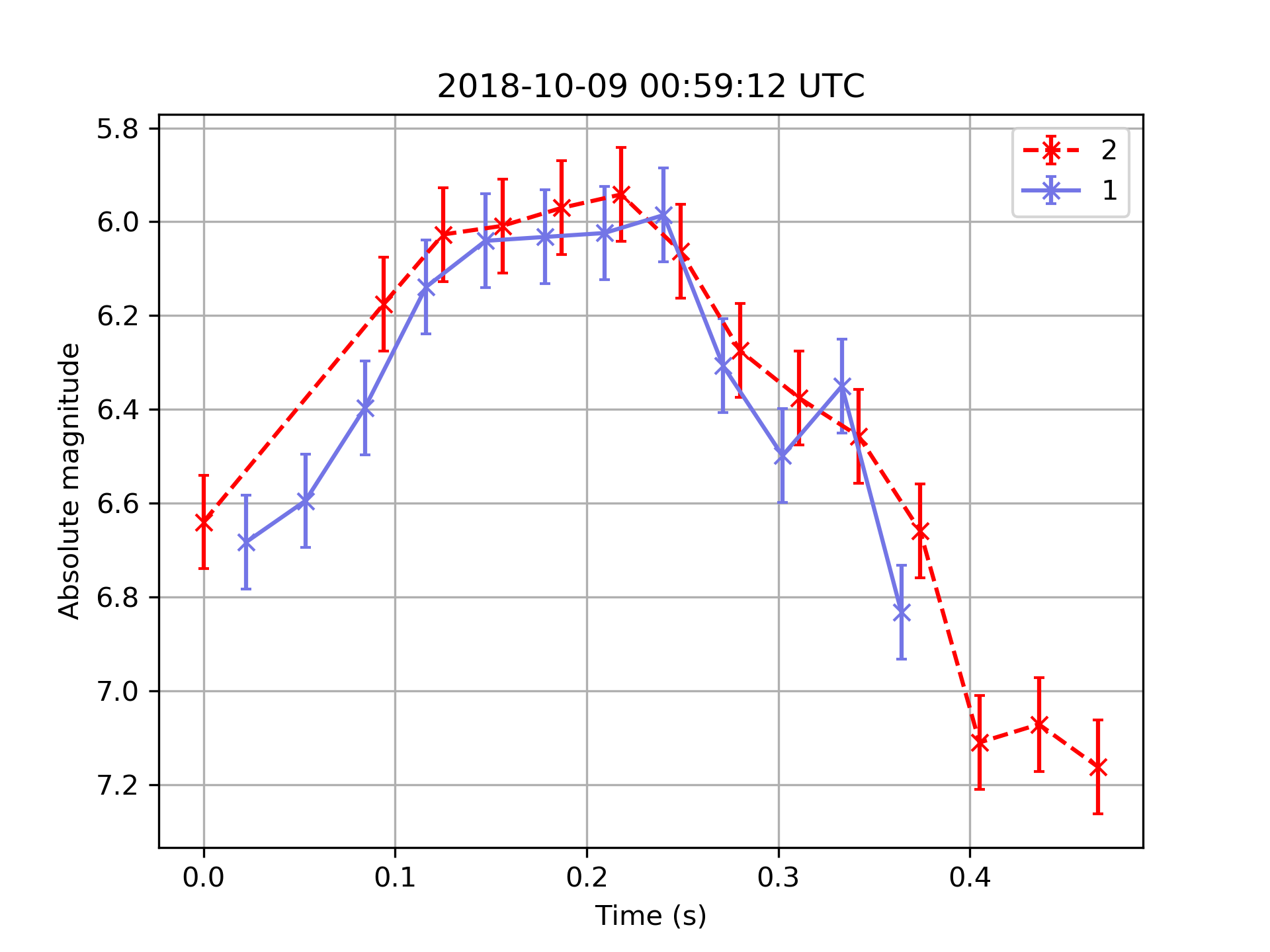}
  \caption{Example light curves of EMCCD Draconids, covering the range of observed peak magnitudes.}
  \label{fig:emccd_lightcurves}
\end{figure*}

\end{appendix}

\begin{appendix}
\section{Table of EMCCD Draconids}

\longtab[1]{
\begin{landscape}
\begin{longtable}{l r r r r r r r r r r r r}
\caption{Radiants, orbits, magnitudes and mass proxies for the observed EMCCD Draconids. Rows below every entry list $1\sigma$ uncertainties.} \label{tab:meteor_list} \\
	\hline\hline 
	
	Date and time (UTC) & $\lambda_{\astrosun}$ & $\alpha_g$ & $\delta_g$ & $v_g$ & $a$ & $e$ & $q$ & $\omega$ & $i$ & $\pi$ & Mag & $\log_{10} I^*$ \\
	            & (deg) & (deg) & (deg) & (\SI{}{\kilo \metre \per \second}) & (AU) &  & (AU) & (deg) & (deg) & (deg) \\
\hline
\endfirsthead
\caption{continued.} \\
	\hline\hline 
	
	Date and time (UTC) & $\lambda_{\astrosun}$ & $\alpha_g$ & $\delta_g$ & $v_g$ & $a$ & $e$ & $q$ & $\omega$ & $i$ & $\pi$ & Mag & $\log_{10} I^*$ \\
	            & (deg) & (deg) & (deg) & (\SI{}{\kilo \metre \per \second}) & (AU) &  & (AU) & (deg) & (deg) & (deg) \\
\hline
\endhead 
\hline
\endfoot
2018-10-09 00:06:37 & 195.4007 & 261.148 & 55.909 & 18.733 & 2.393 & 0.5840 & 0.9954 & 171.91 & 29.37 & 7.31 & 4.19 & -2.20 \\
  &   & 0.043 & 0.025 & 0.064 & 0.020 & 0.0035 & 0.00004 & 0.04 & 0.07 & 0.04 & & \\
2018-10-09 00:07:52 & 195.4016 & 262.200 & 56.273 & 20.328 & 2.993 & 0.6673 & 0.9960 & 172.85 & 31.30 & 8.26 & 4.05 & -2.15 \\
  &   & 0.025 & 0.036 & 0.044 & 0.016 & 0.0020 & 0.00002 & 0.03 & 0.06 & 0.03 & & \\
2018-10-09 00:08:31 & 195.4020 & 262.814 & 55.988 & 21.698 & 4.069 & 0.7552 & 0.9961 & 173.24 & 32.69 & 8.65 & 4.73 & -2.31 \\
  &   & 0.152 & 0.034 & 0.667 & 0.444 & 0.0394 & 0.00001 & 0.12 & 0.73 & 0.12 & & \\
2018-10-09 00:10:08 & 195.4031 & 262.062 & 56.168 & 19.090 & 2.501 & 0.6018 & 0.9960 & 172.68 & 29.84 & 8.08 & 3.32 & -1.86 \\
  &   & 0.018 & 0.016 & 0.043 & 0.014 & 0.0022 & 0.00001 & 0.01 & 0.05 & 0.01 & & \\
2018-10-09 00:10:20 & 195.4033 & 260.831 & 55.913 & 20.383 & 3.024 & 0.6710 & 0.9949 & 171.75 & 31.31 & 7.16 & 3.84 & -1.96 \\
  &   & 0.107 & 0.010 & 0.046 & 0.029 & 0.0030 & 0.00007 & 0.08 & 0.05 & 0.08 & & \\
2018-10-09 00:10:47 & 195.4036 & 262.436 & 55.955 & 20.823 & 3.356 & 0.7032 & 0.9959 & 172.94 & 31.74 & 8.35 & 0.94 & -0.82 \\
  &   & 0.015 & 0.012 & 0.008 & 0.003 & 0.0002 & 0.00001 & 0.01 & 0.01 & 0.01 & & \\
2018-10-09 00:12:51 & 195.4050 & 261.350 & 55.779 & 19.598 & 2.715 & 0.6334 & 0.9953 & 172.06 & 30.34 & 7.46 & 5.04 & -2.36 \\
  &   & 0.018 & 0.022 & 0.032 & 0.011 & 0.0014 & 0.00001 & 0.02 & 0.04 & 0.02 & & \\
2018-10-09 00:14:00 & 195.4058 & 261.414 & 56.034 & 19.739 & 2.735 & 0.6361 & 0.9955 & 172.19 & 30.59 & 7.59 & 5.41 & -2.51 \\
  &   & 0.025 & 0.008 & 0.021 & 0.010 & 0.0013 & 0.00002 & 0.02 & 0.02 & 0.02 & & \\
2018-10-09 00:14:36 & 195.4062 & 262.067 & 55.961 & 19.702 & 2.758 & 0.6389 & 0.9959 & 172.64 & 30.49 & 8.05 & 5.56 & -2.72 \\
  &   & 0.045 & 0.023 & 0.077 & 0.034 & 0.0048 & 0.00002 & 0.03 & 0.08 & 0.03 & & \\
2018-10-09 00:14:54 & 195.4064 & 262.214 & 55.958 & 21.019 & 3.472 & 0.7132 & 0.9958 & 172.79 & 31.97 & 8.20 & 2.77 & -1.62 \\
  &   & 0.015 & 0.024 & 0.023 & 0.010 & 0.0008 & 0.00001 & 0.02 & 0.03 & 0.02 & & \\
2018-10-09 00:17:38 & 195.4083 & 262.062 & 56.143 & 20.165 & 2.932 & 0.6604 & 0.9958 & 172.71 & 31.08 & 8.12 & 3.24 & -1.86 \\
  &   & 0.026 & 0.015 & 0.112 & 0.062 & 0.0067 & 0.00001 & 0.02 & 0.12 & 0.02 & & \\
2018-10-09 00:17:52 & 195.4084 & 262.133 & 55.744 & 21.280 & 3.724 & 0.7326 & 0.9956 & 172.68 & 32.19 & 8.09 & 5.98 & -2.83 \\
  &   & 0.139 & 0.047 & 0.398 & 0.250 & 0.0238 & 0.00003 & 0.10 & 0.43 & 0.10 & & \\
2018-10-09 00:17:54 & 195.4084 & 261.603 & 56.081 & 19.933 & 2.817 & 0.6466 & 0.9956 & 172.35 & 30.82 & 7.76 & 5.12 & -2.36 \\
  &   & 0.042 & 0.007 & 0.031 & 0.015 & 0.0019 & 0.00002 & 0.03 & 0.04 & 0.03 & & \\
2018-10-09 00:18:22 & 195.4088 & 262.639 & 55.737 & 19.682 & 2.808 & 0.6453 & 0.9961 & 172.99 & 30.37 & 8.40 & 1.99 & -1.23 \\
  &   & 0.348 & 0.130 & 0.316 & 0.168 & 0.0191 & 0.00025 & 0.28 & 0.34 & 0.28 & & \\
2018-10-09 00:18:25 & 195.4088 & 261.974 & 56.042 & 19.818 & 2.789 & 0.6430 & 0.9958 & 172.60 & 30.65 & 8.01 & 5.21 & -2.47 \\
  &   & 0.013 & 0.008 & 0.026 & 0.013 & 0.0016 & 0.00001 & 0.01 & 0.03 & 0.01 & & \\
2018-10-09 00:18:44 & 195.4090 & 262.370 & 55.669 & 19.780 & 2.851 & 0.6507 & 0.9959 & 172.78 & 30.47 & 8.19 & 3.58 & -1.95 \\
  &   & 0.013 & 0.026 & 0.029 & 0.009 & 0.0011 & 0.00001 & 0.02 & 0.04 & 0.02 & & \\
2018-10-09 00:19:46 & 195.4097 & 262.432 & 55.913 & 21.927 & 4.300 & 0.7684 & 0.9958 & 172.96 & 32.93 & 8.37 & 1.56 & -1.05 \\
  &   & 0.128 & 0.066 & 0.040 & 0.025 & 0.0013 & 0.00010 & 0.10 & 0.06 & 0.10 & & \\
2018-10-09 00:19:54 & 195.4098 & 260.962 & 56.355 & 19.669 & 2.642 & 0.6233 & 0.9953 & 171.95 & 30.63 & 7.36 & 5.49 & -2.76 \\
  &   & 0.105 & 0.064 & 0.184 & 0.062 & 0.0093 & 0.00009 & 0.10 & 0.23 & 0.10 & & \\
2018-10-09 00:19:56 & 195.4098 & 260.852 & 55.758 & 19.673 & 2.729 & 0.6354 & 0.9950 & 171.69 & 30.44 & 7.10 & 5.64 & -2.57 \\
  &   & 0.388 & 0.141 & 0.040 & 0.013 & 0.0017 & 0.00030 & 0.33 & 0.07 & 0.33 & & \\
2018-10-09 00:20:13 & 195.4100 & 261.106 & 55.483 & 21.299 & 3.739 & 0.7339 & 0.9948 & 171.87 & 32.17 & 7.28 & 3.59 & -1.91 \\
  &   & 0.093 & 0.034 & 0.022 & 0.019 & 0.0013 & 0.00007 & 0.08 & 0.02 & 0.08 & & \\
2018-10-09 00:24:15 & 195.4128 & 261.709 & 56.185 & 19.394 & 2.592 & 0.6159 & 0.9957 & 172.43 & 30.22 & 7.85 & 5.50 & -2.72 \\
  &   & 0.041 & 0.041 & 0.134 & 0.058 & 0.0081 & 0.00003 & 0.03 & 0.14 & 0.03 & & \\
2018-10-09 00:24:52 & 195.4132 & 261.638 & 55.836 & 20.452 & 3.118 & 0.6807 & 0.9954 & 172.32 & 31.32 & 7.73 & 5.54 & -2.55 \\
  &   & 0.018 & 0.024 & 0.045 & 0.022 & 0.0023 & 0.00001 & 0.02 & 0.05 & 0.02 & & \\
2018-10-09 00:25:23 & 195.4136 & 262.628 & 56.100 & 21.198 & 3.589 & 0.7225 & 0.9961 & 173.13 & 32.20 & 8.55 & 3.09 & -1.68 \\
  &   & 0.034 & 0.004 & 0.023 & 0.019 & 0.0015 & 0.00002 & 0.03 & 0.02 & 0.03 & & \\
2018-10-09 00:25:59 & 195.4140 & 262.163 & 56.083 & 20.088 & 2.911 & 0.6579 & 0.9959 & 172.76 & 30.97 & 8.17 & 3.32 & -1.83 \\
  &   & 0.009 & 0.014 & 0.017 & 0.006 & 0.0007 & 0.00001 & 0.01 & 0.02 & 0.01 & & \\
2018-10-09 00:26:48 & 195.4145 & 262.170 & 56.165 & 19.224 & 2.550 & 0.6095 & 0.9960 & 172.76 & 29.99 & 8.18 & 5.53 & -2.67 \\
  &   & 0.017 & 0.019 & 0.038 & 0.010 & 0.0016 & 0.00001 & 0.02 & 0.05 & 0.02 & & \\
2018-10-09 00:29:09 & 195.4162 & 262.108 & 56.096 & 19.895 & 2.819 & 0.6467 & 0.9959 & 172.72 & 30.76 & 8.13 & 2.31 & -1.37 \\
  &   & 0.012 & 0.010 & 0.009 & 0.003 & 0.0003 & 0.00001 & 0.01 & 0.01 & 0.01 & & \\
2018-10-09 00:30:10 & 195.4169 & 262.123 & 55.988 & 20.920 & 3.391 & 0.7064 & 0.9957 & 172.73 & 31.88 & 8.15 & 2.50 & -1.26 \\
  &   & 0.014 & 0.007 & 0.051 & 0.035 & 0.0031 & 0.00001 & 0.01 & 0.05 & 0.01 & & \\
2018-10-09 00:30:24 & 195.4170 & 262.482 & 55.895 & 20.043 & 2.938 & 0.6610 & 0.9960 & 172.93 & 30.84 & 8.35 & 4.51 & -2.34 \\
  &   & 0.125 & 0.086 & 0.096 & 0.034 & 0.0038 & 0.00010 & 0.11 & 0.13 & 0.11 & & \\
2018-10-09 00:30:45 & 195.4173 & 262.300 & 56.127 & 19.325 & 2.596 & 0.6162 & 0.9961 & 172.85 & 30.10 & 8.27 & 4.33 & -2.33 \\
  &   & 0.015 & 0.032 & 0.029 & 0.006 & 0.0009 & 0.00001 & 0.02 & 0.04 & 0.02 & & \\
2018-10-09 00:30:57 & 195.4174 & 263.270 & 55.346 & 19.881 & 2.997 & 0.6676 & 0.9964 & 173.34 & 30.45 & 8.76 & 4.93 & -2.47 \\
  &   & 0.232 & 0.136 & 0.109 & 0.035 & 0.0034 & 0.00017 & 0.18 & 0.16 & 0.18 & & \\
2018-10-09 00:33:05 & 195.4189 & 262.340 & 55.534 & 20.073 & 3.015 & 0.6697 & 0.9958 & 172.72 & 30.77 & 8.14 & 3.04 & -1.46 \\
  &   & 0.014 & 0.053 & 0.065 & 0.026 & 0.0028 & 0.00001 & 0.02 & 0.09 & 0.02 & & \\
2018-10-09 00:33:54 & 195.4194 & 261.699 & 56.214 & 19.620 & 2.672 & 0.6273 & 0.9957 & 172.44 & 30.49 & 7.86 & 3.68 & -1.94 \\
  &   & 0.014 & 0.018 & 0.033 & 0.011 & 0.0016 & 0.00001 & 0.01 & 0.04 & 0.01 & & \\
2018-10-09 00:34:12 & 195.4196 & 261.985 & 55.767 & 20.477 & 3.165 & 0.6854 & 0.9956 & 172.55 & 31.31 & 7.97 & 2.88 & -1.61 \\
  &   & 0.021 & 0.029 & 0.036 & 0.017 & 0.0017 & 0.00001 & 0.01 & 0.05 & 0.01 & & \\
2018-10-09 00:34:35 & 195.4199 & 261.820 & 55.782 & 20.460 & 3.143 & 0.6832 & 0.9955 & 172.43 & 31.31 & 7.85 & 4.71 & -2.49 \\
  &   & 0.025 & 0.091 & 0.080 & 0.025 & 0.0025 & 0.00002 & 0.03 & 0.12 & 0.03 & & \\
2018-10-09 00:34:57 & 195.4201 & 261.592 & 56.010 & 18.759 & 2.401 & 0.5854 & 0.9957 & 172.27 & 29.42 & 7.69 & 5.19 & -2.52 \\
  &   & 0.018 & 0.034 & 0.036 & 0.008 & 0.0014 & 0.00002 & 0.02 & 0.05 & 0.02 & & \\
2018-10-09 00:40:00 & 195.4236 & 262.710 & 54.921 & 17.099 & 2.105 & 0.5266 & 0.9963 & 172.76 & 27.04 & 8.18 & 5.63 & -2.83 \\
  &   & 0.193 & 0.127 & 0.370 & 0.102 & 0.0183 & 0.00012 & 0.15 & 0.46 & 0.15 & & \\
2018-10-09 00:40:02 & 195.4236 & 262.162 & 56.212 & 20.711 & 3.210 & 0.6897 & 0.9959 & 172.82 & 31.72 & 8.24 & 4.69 & -2.35 \\
  &   & 0.041 & 0.011 & 0.027 & 0.017 & 0.0016 & 0.00002 & 0.03 & 0.03 & 0.03 & & \\
2018-10-09 00:42:09 & 195.4251 & 261.976 & 55.944 & 19.376 & 2.629 & 0.6211 & 0.9958 & 172.55 & 30.11 & 7.98 & 4.78 & -2.45 \\
  &   & 0.262 & 0.136 & 0.059 & 0.010 & 0.0016 & 0.00020 & 0.22 & 0.10 & 0.22 & & \\
2018-10-09 00:42:24 & 195.4252 & 261.600 & 55.887 & 19.625 & 2.719 & 0.6339 & 0.9955 & 172.27 & 30.40 & 7.70 & 4.43 & -2.30 \\
  &   & 0.010 & 0.013 & 0.022 & 0.007 & 0.0010 & 0.00001 & 0.01 & 0.03 & 0.01 & & \\
2018-10-09 00:43:13 & 195.4258 & 261.918 & 56.138 & 20.173 & 2.930 & 0.6601 & 0.9957 & 172.60 & 31.10 & 8.03 & 4.36 & -2.05 \\
  &   & 0.014 & 0.019 & 0.020 & 0.006 & 0.0006 & 0.00001 & 0.01 & 0.03 & 0.01 & & \\
2018-10-09 00:44:30 & 195.4267 & 262.635 & 56.235 & 21.228 & 3.572 & 0.7211 & 0.9961 & 173.18 & 32.28 & 8.60 & 2.06 & -1.33 \\
  &   & 0.281 & 0.125 & 0.017 & 0.023 & 0.0018 & 0.00022 & 0.24 & 0.04 & 0.24 & & \\
2018-10-09 00:45:10 & 195.4272 & 261.695 & 56.232 & 21.624 & 3.816 & 0.7391 & 0.9954 & 172.52 & 32.76 & 7.95 & 3.65 & -1.80 \\
  &   & 0.032 & 0.043 & 0.058 & 0.035 & 0.0026 & 0.00003 & 0.04 & 0.08 & 0.04 & & \\
2018-10-09 00:46:20 & 195.4280 & 262.032 & 56.029 & 20.632 & 3.198 & 0.6886 & 0.9957 & 172.66 & 31.57 & 8.09 & 4.19 & -2.12 \\
  &   & 0.030 & 0.030 & 0.038 & 0.015 & 0.0014 & 0.00002 & 0.02 & 0.05 & 0.02 & & \\
2018-10-09 00:48:59 & 195.4298 & 262.503 & 56.053 & 20.324 & 3.048 & 0.6733 & 0.9961 & 173.00 & 31.21 & 8.43 & 4.49 & -2.09 \\
  &   & 0.143 & 0.068 & 0.016 & 0.009 & 0.0010 & 0.00011 & 0.12 & 0.03 & 0.12 & & \\
2018-10-09 00:49:03 & 195.4298 & 262.424 & 56.205 & 22.928 & 5.589 & 0.8218 & 0.9958 & 173.07 & 34.10 & 8.50 & 3.45 & -1.57 \\
  &   & 0.034 & 0.008 & 0.105 & 0.173 & 0.0067 & 0.00001 & 0.03 & 0.11 & 0.03 & & \\
2018-10-09 00:50:26 & 195.4308 & 260.164 & 56.377 & 18.070 & 2.146 & 0.5364 & 0.9950 & 171.25 & 28.75 & 6.69 & 6.31 & -3.11 \\
  &   & 0.089 & 0.040 & 0.102 & 0.031 & 0.0062 & 0.00005 & 0.07 & 0.11 & 0.07 & & \\
2018-10-09 00:50:30 & 195.4308 & 261.233 & 56.480 & 18.889 & 2.376 & 0.5810 & 0.9956 & 172.14 & 29.73 & 7.57 & 4.02 & -2.12 \\
  &   & 0.016 & 0.013 & 0.013 & 0.003 & 0.0005 & 0.00001 & 0.01 & 0.02 & 0.01 & & \\
2018-10-09 00:53:49 & 195.4331 & 262.207 & 56.102 & 20.183 & 2.954 & 0.6629 & 0.9959 & 172.80 & 31.08 & 8.23 & 4.53 & -2.20 \\
  &   & 0.026 & 0.007 & 0.017 & 0.010 & 0.0011 & 0.00001 & 0.02 & 0.02 & 0.02 & & \\
2018-10-09 00:54:02 & 195.4332 & 262.175 & 56.064 & 19.528 & 2.676 & 0.6278 & 0.9960 & 172.74 & 30.32 & 8.17 & 6.61 & -3.38 \\
  &   & 0.010 & 0.020 & 0.020 & 0.005 & 0.0006 & 0.00001 & 0.01 & 0.03 & 0.01 & & \\
2018-10-09 00:54:33 & 195.4336 & 262.130 & 55.991 & 18.750 & 2.416 & 0.5878 & 0.9960 & 172.66 & 29.38 & 8.09 & 3.32 & -1.72 \\
  &   & 0.010 & 0.019 & 0.015 & 0.002 & 0.0004 & 0.00001 & 0.01 & 0.02 & 0.01 & & \\
2018-10-09 00:59:05 & 195.4367 & 262.004 & 56.005 & 21.422 & 3.746 & 0.7342 & 0.9956 & 172.66 & 32.44 & 8.10 & 1.05 & -0.76 \\
  &   & 0.029 & 0.008 & 0.041 & 0.036 & 0.0024 & 0.00002 & 0.02 & 0.04 & 0.02 & & \\
2018-10-09 00:59:12 & 195.4368 & 261.853 & 56.040 & 19.636 & 2.710 & 0.6325 & 0.9957 & 172.50 & 30.45 & 7.94 & 5.94 & -2.86 \\
  &   & 0.054 & 0.068 & 0.070 & 0.019 & 0.0026 & 0.00005 & 0.06 & 0.10 & 0.06 & & \\
2018-10-09 00:59:30 & 195.4370 & 260.516 & 55.773 & 18.862 & 2.428 & 0.5902 & 0.9949 & 171.39 & 29.51 & 6.83 & 6.04 & -2.94 \\
  &   & 0.069 & 0.014 & 0.024 & 0.009 & 0.0015 & 0.00005 & 0.06 & 0.03 & 0.06 & & \\
2018-10-09 00:59:35 & 195.4370 & 261.633 & 56.017 & 19.653 & 2.712 & 0.6328 & 0.9956 & 172.33 & 30.47 & 7.77 & 3.46 & -1.62 \\
  &   & 0.009 & 0.010 & 0.013 & 0.004 & 0.0005 & 0.00001 & 0.01 & 0.02 & 0.01 & & \\
2018-10-09 00:59:50 & 195.4372 & 261.715 & 55.772 & 19.379 & 2.644 & 0.6235 & 0.9956 & 172.31 & 30.07 & 7.75 & 2.37 & -1.28 \\
  &   & 0.014 & 0.019 & 0.015 & 0.003 & 0.0005 & 0.00001 & 0.01 & 0.02 & 0.01 & & \\
2018-10-09 01:01:43 & 195.4385 & 261.983 & 56.221 & 20.594 & 3.132 & 0.6820 & 0.9958 & 172.68 & 31.60 & 8.12 & 6.04 & -2.97 \\
  &   & 0.117 & 0.047 & 0.077 & 0.039 & 0.0049 & 0.00008 & 0.10 & 0.08 & 0.10 & & \\
2018-10-09 01:08:52 & 195.4434 & 262.068 & 55.950 & 20.164 & 2.967 & 0.6643 & 0.9958 & 172.65 & 31.02 & 8.09 & 5.31 & -2.59 \\
  &   & 0.174 & 0.042 & 0.501 & 0.331 & 0.0298 & 0.00004 & 0.13 & 0.54 & 0.13 & & \\
2018-10-09 01:10:16 & 195.4444 & 263.413 & 56.213 & 21.256 & 3.654 & 0.7273 & 0.9965 & 173.71 & 32.27 & 9.16 & 4.22 & -2.13 \\
  &   & 0.088 & 0.017 & 0.047 & 0.033 & 0.0028 & 0.00005 & 0.07 & 0.05 & 0.07 & & \\
2018-10-09 01:12:26 & 195.4458 & 261.633 & 55.913 & 19.390 & 2.625 & 0.6208 & 0.9956 & 172.29 & 30.13 & 7.74 & 4.80 & -2.38 \\
  &   & 0.043 & 0.012 & 0.022 & 0.009 & 0.0013 & 0.00003 & 0.04 & 0.03 & 0.04 & & \\
2018-10-09 01:13:18 & 195.4464 & 261.579 & 55.956 & 19.086 & 2.511 & 0.6035 & 0.9956 & 172.25 & 29.79 & 7.70 & 5.14 & -2.54 \\
  &   & 0.261 & 0.155 & 0.102 & 0.040 & 0.0062 & 0.00023 & 0.24 & 0.12 & 0.24 & & \\
2018-10-09 01:14:12 & 195.4471 & 261.196 & 56.256 & 19.152 & 2.483 & 0.5990 & 0.9955 & 172.06 & 29.98 & 7.50 & 4.57 & -2.28 \\
  &   & 0.019 & 0.021 & 0.013 & 0.002 & 0.0004 & 0.00001 & 0.02 & 0.02 & 0.02 & & \\
2018-10-09 01:14:40 & 195.4474 & 261.443 & 56.452 & 20.109 & 2.823 & 0.6474 & 0.9956 & 172.34 & 31.15 & 7.79 & 5.30 & -2.63 \\
  &   & 0.080 & 0.060 & 0.039 & 0.011 & 0.0014 & 0.00007 & 0.07 & 0.06 & 0.07 & & \\
2018-10-09 01:15:32 & 195.4480 & 261.452 & 55.902 & 20.320 & 3.024 & 0.6709 & 0.9953 & 172.19 & 31.21 & 7.64 & 3.44 & -1.92 \\
  &   & 0.139 & 0.086 & 0.009 & 0.012 & 0.0013 & 0.00013 & 0.13 & 0.02 & 0.13 & & \\
2018-10-09 01:17:31 & 195.4493 & 261.989 & 56.232 & 20.535 & 3.097 & 0.6785 & 0.9958 & 172.69 & 31.54 & 8.14 & 4.19 & -1.99 \\
  &   & 0.106 & 0.062 & 0.018 & 0.013 & 0.0013 & 0.00009 & 0.10 & 0.03 & 0.10 & & \\
2018-10-09 01:21:22 & 195.4520 & 261.745 & 55.998 & 19.711 & 2.742 & 0.6369 & 0.9956 & 172.41 & 30.53 & 7.86 & 4.64 & -2.35 \\
  &   & 0.101 & 0.062 & 0.017 & 0.010 & 0.0013 & 0.00009 & 0.09 & 0.02 & 0.09 & & \\
2018-10-09 01:21:30 & 195.4521 & 262.764 & 56.139 & 20.150 & 2.956 & 0.6630 & 0.9963 & 173.20 & 31.03 & 8.66 & 5.20 & -2.69 \\
  &   & 0.145 & 0.021 & 0.168 & 0.066 & 0.0094 & 0.00008 & 0.12 & 0.19 & 0.12 & & \\
2018-10-09 01:23:21 & 195.4533 & 262.104 & 55.876 & 20.780 & 3.327 & 0.7007 & 0.9957 & 172.67 & 31.69 & 8.12 & 1.82 & -1.19 \\
  &   & 0.026 & 0.004 & 0.014 & 0.009 & 0.0008 & 0.00002 & 0.02 & 0.02 & 0.02 & & \\
2018-10-09 01:27:34 & 195.4562 & 261.840 & 55.849 & 20.001 & 2.897 & 0.6563 & 0.9956 & 172.44 & 30.81 & 7.90 & 4.11 & -2.25 \\
  &   & 0.057 & 0.083 & 0.121 & 0.061 & 0.0077 & 0.00003 & 0.04 & 0.13 & 0.04 & & \\
\end{longtable}
\end{landscape}
}

\end{appendix}

\end{document}